\documentclass[preprint,showpacs,preprintnumbers,superscriptaddress]{revtex4-1}
\usepackage{amsmath,amssymb,graphicx,bm,subfigure,float,siunitx,color}
\usepackage{natbib}
\usepackage{bm,subfigure,float,siunitx,graphicx}
\usepackage[figuresright]{rotating}
\usepackage{color}
\usepackage{epstopdf}
\usepackage{mathrsfs}
\newcommand{\be}{\begin{equation}}
\newcommand{\ee}{\end{equation}}

\newcommand{\1}{\left}
\newcommand{\2}{\right}
\def\({\left(}
\def\){\right)}
\def\[{\left[}
\def\]{\right]}

\newcommand{\dif}{\,\mathrm{d}}

\newcommand{\p}{\partial}

\newcommand{\m}{\mu}
\newcommand{\n}{\nu}
\newcommand{\al}{\alpha}
\newcommand{\bet}{\beta}

\newcommand{\na}{\nabla}

\begin{document}
\title{\boldmath Newtonian potential from scattering amplitudes in super-renormalizable gravity}

\author{Haiyuan Feng\footnote{Corresponding author}}
\email{Email address: fenghaiyuan@sxnu.edu.cn }
\affiliation{School of Physics and Electronic Engineering, Shanxi Normal University, Taiyuan 030031, China}

\author{Rong-Jia Yang\footnote{Corresponding author}}
\email{Email address: yangrongjia@tsinghua.org.cn}
\affiliation{College of Physical Science and Technology, Hebei University, Baoding 071002, China}

\author{Jinjun Zhang\footnote{Corresponding author}}
\email{Email address: zhangjinjun@sxnu.edu.cn }
\affiliation{School of Physics and Electronic Engineering, Shanxi Normal University, Taiyuan 030031, China}

\begin{abstract}
Based on the classical limit of relativistic scattering amplitudes, we compute the coupling between a general super-renormalizable gravity and massive scalar particles. This allows us to derive the $D$-dimensional metric corrections at both tree-level and one-loop level—the latter containing the first calculation by using newly derived three-graviton Feynman rules. By introducing Newtonian potential function $\Phi\equiv\frac{h_{00}}{2}$, we reproduce at tree level, $\mathcal{O}(G)$, the well-known result proportional to the error function in four-dimensional spacetime. Furthermore, we obtain the loop-level contributions to the potential function at order $\mathcal{O}(G^2)$ and perform a numerical analysis of its behavior at large distances. Our results indicate that, at the one-loop level, the magnitude of the potential increases gradually with distance in the asymptotic regime.
\end{abstract}

\maketitle

\section{Introduction}
The challenge of formulating a consistent theory of quantum gravity (QG) has persisted for decades, with various approaches proposed in an effort to reconcile quantum mechanics with general relativity (GR). One of the central concerns in QG is the issue of renormalizability, particularly in the ultraviolet (UV) regime. This is a critical region, as it is where both quantum and classical singularities manifest. Unfortunately, GR proves problematic in the UV limit, suffering from divergences that render it non-renormalizable \cite{Utiyama:1962sn,Buchbinder:1992rb,Shapiro:2008sf}. However, effective quantum gravity in the classical limit offers a viable framework for recovering GR, where quantum field theory (QFT) techniques are employed to derive classical results from a quantum perspective \cite{Duff:1973zz,Bjerrum-Bohr:2018xdl,Goldberger:2004jt,Kosower:2018adc,Cheung:2018wkq,Bern:2019nnu,Bern:2019crd,Cristofoli:2019neg,Chung:2019yfs}. Within this framework, gravitational interactions are treated as mediated by spin-2 gravitons, and GR is recast in the formalism of QFT \cite{Donoghue:1995cz}. In addition, the analytical techniques developed in QFT are crucial for refining theoretical predictions, particularly in the study of gravitational waves. In this context, modern approaches such as on-shell scattering amplitudes and generalized unitarity have proven especially effective. While quantum corrections to gravity remain negligible in current experimental settings, they continue to be of great interest in the ongoing research into QG. These methods offer new insights into gravitational phenomena, potentially revealing deeper connections between quantum mechanics and spacetime dynamics.

Scattering amplitudes provide a uniquely powerful framework for probing QG, with gauge-invariant observables laying the foundation for a future QG phenomenology \cite{Kalin:2020fhe,Bern:2021yeh,Bern:2021dqo,Adamo:2022qci,Bohnenblust:2024hkw,Jakobsen:2021lvp,Mougiakakos:2024nku}. Their theoretical precision enables the systematic extraction of QG effects, including the constraining of higher-derivative operators induced by loop divergences in effective approaches \cite{Wang:2015jna}.
It is well-established that even at the tree-level, intermediate computational steps often display a complex structure that typically cancels out in the final physical amplitudes \cite{DeWitt:1967uc,Berends:1974gk,Elvang:2013cua}. This complexity arises from the inherent ambiguity of the off-shell Lagrangian formalism, which is invariant under gauge transformations and field redefinitions. Recently, there has been a surge of interest in developing advanced methods for simplifying the calculation of scattering amplitudes. 
The primary methods employed in this context are unitarity techniques, which isolate the "cut-constructible" components of loop amplitudes, specifically targeting the non-analytic terms \cite{Bern:1994zx}. The technique has significantly streamlined amplitude calculations, making many of the current computations for the Standard Model in particle physics feasible, where they would otherwise have been intractable. In classical gravity, the long-range contributions of interest are precisely of this non-analytic form, depending on the dimensionless ratio $\frac{m}{\sqrt{-q^2}}$, where $m$ represents a massive probe and $q$ is the appropriately defined momentum \cite{PhysRevLett.72.2996}. This observation has prompted the suggestion that these modern methods be employed to compute post-Newtonian and post-Minkowskian corrections within GR, particularly in the context of astrophysical phenomena such as binary black hole mergers \cite{Blanchet:2006jqj,Fraser:2014aga,Minazzoli:2009zz,LIGOScientific:2016aoc}. The application of these advanced computational tools is expected to enhance the precision of theoretical models and provide valuable insights into gravitational wave physics and other relativistic phenomena.


In addition, super-renormalizable gravity offers a consistent and predictive framework for ultraviolet-complete QG, preserving unitarity, covariance, and perturbative control across all energy scales \cite{Biswas:2011ar, Modesto:2011kw, Briscese:2013lna, Tomboulis:2015esa, Calcagni:2014vxa, Alexander:2012aw, Dona:2015tra}. A defining feature of this approach is weak nonlocality—a property shared by string theory \cite{Brekke:1987ptq}—which ensures super-renormalizability while effectively eliminating pathological states. 
While traditional perturbative studies in super-renormalizable gravity have successfully established a finite Newtonian potential at $\mathcal{O}(G)$ \cite{Biswas:2011ar} and explored various phenomenological implications \cite{Briscese:2013lna, Krasnikov:1987yj, Modesto:2011kw, Tomboulis:2015esa, Dona:2015tra}, the structure of loop-induced metric corrections within this class of theories has remained far less explored. In this work, we address this gap by employing modern scattering-amplitude techniques to investigate the classical limit of the relativistic amplitude at one-loop order, thereby providing a systematic analysis of the leading long-range $\mathcal{O}(G^2)$ contributions to the effective static gravitational interaction.

The paper is organized as follows: In Section II, we introduce the 
super-renormalizable gravity from the perspectives of unitarity and power counting. In Section III, we give the relation between the scattering amplitudes and the matter fields in the process of gravitational radiation from a massive scalar particle, and subsequently obtained the general expression for the higher-order metric correction. Using this method, we calculate the metric correction and Newtonian potential in $D$-dimensional spacetime at the tree-level. In Section IV, by constructing the Feynman rules for the three-graviton vertex, we investigate the loop-level correction to the Newtonian potential. Furthermore, we numerically analyze the behavior of the potential function as a function of distance in the large-distance regime. In Section V, where we also draw our conclusions. 

\section{Super-renormalizable Gravity}
Due to the intrinsic limitations of GR, a variety of modified gravity models have emerged naturally over time as attempts to extend or replace the GR, each aiming to address unresolved issues such as the nature of dark energy, singularities, and the behavior of gravity at small scales. These models include a range of modifications to the Einstein-Hilbert action, from higher-order derivatives of the curvature tensor to nonlocal corrections that introduce additional degrees of freedom (d.o.f) to the gravitational field \cite{Nojiri:2017ncd,Nojiri:2010wj,Nojiri:2006ri,Capozziello:2011et,Faraoni:2010pgm,delaCruz-Dombriz:2012bni,Olmo:2011uz}. Among these modifications, string-inspired nonlocal gravity has attracted significant attention due to their promising quantum properties and their potential to provide a more robust framework for understanding gravity at quantum scales \cite{Deser:2007jk}. Initially, the nonlocal gravity was primarily proposed as a possible explanation for the accelerated expansion of the universe. However, the model quickly evolved to provide a broader framework for addressing a variety of quantum gravity phenomena \cite{ Bamba:2012ky, Zhang:2011uv, Elizalde:2011su, Nojiri:2010pw}. The profound impact of nonlocal gravity is particularly evident in their connection to string theory, where nonlocal terms are commonly introduced to describe string interactions at high energies. Additionally, the model offers a fresh perspective on the mathematical structure of gravity, with modifications to the gravitational action that go beyond traditional local interaction, introducing both theoretical elegance and physical insights into the UV regime \cite{Nojiri:2007uq, Zhang:2016ykx}.

To describe physical phenomena at both classical and quantum levels, most nonlocal models introduce modifications involving either nonlocal scalar fields or terms that involve the d'Alembertian operator $\Box$. These terms allow for a nonlocal dependence of the gravitational field on spacetime, in contrast to the local interactions present in GR. Without losing generality, we focus here on a general nonlocal model described by the following action
\be
\label{1}
S_{0}=\frac{2}{\kappa^2}\int \sqrt{-g}\dif^{D}x\(R+RF_0(\Box)R+R_{\m\n}F_{2}(\Box)R^{\m\n}+R_{\m\n\al\bet}F_{4}(\Box)R^{\m\n\al\bet}+V\),
\ee
where $R$ is the Riemann curvature scalar, $\kappa^2=32\pi G_{N}$, and a set of local terms $V$ cubic or higher in curvature. The latter involves operators with a carefully selected number of derivatives, ensuring that the favorable quantum properties of the theory are preserved. Furthermore, the crucial elements of our analysis are the functions of the covariant d'Alembertian operator $F_i(\Box)$ which called form factors. The form factors are assumed to be entire functions, allowing them to be expanded in a Taylor series  $F_{i}(\Box)=\sum^{\infty}_{n=0}f_{in}\frac{\Box^n}{ M^{2n}_{\ast}}$. where $M_{\ast}$ denotes the mass scale at which the higher-derivative terms in the action become relevant. Therefore, nonlocal gravity can be viewed as an infinite-order derivative extension of higher-derivative gravity. Such a model exhibits massive ghost particle excitations at the Planck scale $M_P$ and induces inflation through scalar particle excitations at the scale $m$ \cite{Stelle:1976gc,Koshelev:2016xqb,Planck:2015fie,Planck:2015sxf}. Consequently, the range for $M_{\ast}$ is naturally chosen as $m < M_{\ast} < M_P$. This phenomenological hierarchy suggests that the microscopical origins of inflation and nonlocal UV-completion of gravity are not the same.

Furthermore, the requirements of super-renormalizability and unitarity constrain the admissible form factors to the following categories
\be
\1\{\begin{split}
\label{2}
&F_{0}(\Box)=-\frac{(D-2)\(e^{H_{0}(\Box)}-1\)+D\(e^{H_{2}(\Box)}-1\)}{4(D-1)\Box}+F_{4}(\Box),\\
&F_{2}(\Box)=\frac{e^{H_{2}(\Box)}-1}{\Box}-4F_{4}(\Box),
\end{split}\2.
\ee
where $F_{4}(\Box)$ may in principle remain arbitrary, super-renormalizability mandates that $F_4(\Box)$ shares the same UV asymptotic behavior as the other two form factors $F_i(\Box)$ ($i = 0, 2$).
To achieve this, the most straightforward approach is to retain only two of the three form factors, which leads to the condition $F_4(\Box) = 0$. Additionally, the entire function $e^{H_i(\Box)}$ must satisfy three distinct sets of criteria \cite{Tomboulis:1997gg,Modesto:2013ioa}:

$\bullet$ The function $ e^{H_{i}(\Box)}$ must be real and positive along the real axis and have no zeros within the entire complex plane for $ z < \infty$ $\(z\equiv-\frac{\Box}{M^2_{\ast}}\)$. This requirement guarantees the absence of gauge-invariant poles, except for the transverse massless physical graviton pole.

$\bullet$ $e^{H_{i}(\Box)}$ exhibits the same asymptotic behavior along the real axis at $\pm\infty$.

$\bullet$ There exist value $ 0<\varphi <\frac{\pi}{2} $, and a positive integer $\gamma $, such that asymptotically
\be
\label{3}
|e^{H_{i}(\Box)}|\rightarrow|z|^{\gamma+N+1},\quad |z|\rightarrow\infty, \quad\text{with}\quad\gamma\geqslant \frac{D_{\text{even}}}{2} \quad\text{or}  \quad \gamma\geqslant \frac{D_{\text{odd}}-1}{2},
\ee
and regin $C$ defined by
\be
\label{4}
C\equiv\{z:-\varphi<\text{arg} z<+\varphi,\pi-\varphi<\text{arg} z<\pi+\varphi\}.
\ee

The final condition is essential for achieving the maximum convergence of the theory in the UV regime. The required asymptotic behavior must be enforced not only along the real axis but also in the conical regions surrounding it. In Euclidean spacetime, condition (ii) is not strictly necessary if condition (iii) is satisfied. In Eq.\eqref{3}, the capital $N$ is defined as a function of the spacetime dimension $D$, with $2N+4=D_{\text{odd}}+1$ for odd dimensions and $2N+4=D_{\text{even}}$ for even dimensions. Moreover, Ref.\cite{Tomboulis:1997gg} has already provided a specific example of a special form factor that satisfies the above conditions, which can be expressed as
\be
\label{5}
e^{H_{i}(\Box)}=e^{\frac{1}{2}\[\Gamma(0,p_{i}(z)^2)+\gamma_{E}+\log\(p_{i}(z)^2\) \]},
\ee
where $\gamma_E \approx 0.577216$ denotes the Euler-Mascheroni constant, $\Gamma(0, z) = \int_{z}^{\infty} \frac{e^{-t}}{t} \dif t$ represents the incomplete Gamma function with its first argument set to zero. The polynomial  $p_i(z)$, which has a degree $\gamma +N+ 1$ and satisfying $p_i(0) = 0$, ensures that the low-energy limit of nonlocal theory is correct. In the UV regime, where $ |z| \gg 1$, the function exhibits polynomial behavior of the form $|z|^{\gamma + N + 1} $ within the conical region around the real axis, with an angular opening given by $\varphi= \frac{\pi}{4(\gamma + N + 1)}$. To achieve super-renormalizability, it is necessary for the degrees of the polynomials defining $H_0(\Box)$ and $H_2(\Box)$ to be identical. Moreover, for simplicity, we will exclude the contributions from the interaction term $V$ in the subsequent analysis.

\subsection{Propagator and Unitarity in Minkowski Spacetime}
To obtain the explicit form of the propagator, we expand the metric tensor around Minkowski spacetime, denoted as $g_{\m\n}=\eta_{\m\n}+h_{\m\n}$. In addition, we consider the following gauge-fixing term
\be
\label{6}
S_{\text{gf}}=\int \dif^{D}x \frac{\eta^{\m\n}G_{\m}\omega(\Box)G_{\n}}{\kappa^2\xi},
\ee
where $G_{\m}$ is taken as $\partial^{\n}h_{\m\n}$ to represent the usual harmonic gauge, $\xi$ denotes the gauge-fixing parameter, and $\omega(\Box)$ is a weight functional \cite{Accioly:2002tz,Buchbinder:1992rb,Stelle:1976gc}. Subsequently, linearizing the action leads to the following expression
\be
\label{7}
S_{0}+S_{gf}\approx\frac{1}{2}\int \frac{\dif^D q}{(2\pi)^D}\tilde{h}^{\m\n}(q)O_{\m\n\al\bet}\tilde{h}^{\al\bet}(-q),\\
\ee
here we have already used the rescaled graviton field $\tilde{h}^{\mu\nu} \equiv \frac{h^{\mu\nu}}{\kappa}$. The operator $O$ is made out of two terms, one coming from the quadratization of Eq.\eqref{1} and the other from the gauge-fixing term. The d'Alembertian operator in $S_0$  and the gauge-fixing term are defined in flat spacetime. After inverting the operator $O$ \cite{Accioly:2002tz} and applying the form factors from Eq.\eqref{2}, we derive the two-point function  in momentum space,
\be
\label{8}
\begin{split}
O^{-1}_{\m\n\al\bet}&=\frac{1}{q^2e^{H_2(-q^2)}} I_{\m\n\al\bet}+\( \frac{2\xi}{q^2\omega(-q^2)}-\frac{2}{q^2 e^{H_2(-q^2)}}  \)J_{\m\n\al\bet}\\
&-\(  \frac{4}{(D-1)(D-2)q^2e^{H_0(-q^2)}}+\frac{4}{(D-1)q^2 e^{H_2(-q^2)} }  \)T_{\m\n\al\bet}\\
&+\( \frac{2}{(D-1)q^2 e^{H_2(-q^2)}}-\frac{2}{(D-1)(D-2)q^2 e^{H_0(-q^2)}} \)C_{\m\n\al\bet}  \\
&+\(\frac{D-2}{(D-1)q^2e^{H_2(-q^2)}}+\frac{1}{(D-2)(D-1)q^2e^{H_0(-q^2)}}-\frac{3\xi}{2q^2\omega(-q^2)} \)K_{\m\n\al\bet},
\end{split}
\ee
where the five independent operators are $I_{\alpha \beta}^{\mu \nu}=\frac{1}{2}\left(\delta_\alpha^\mu \delta_\beta^\nu+\delta_\beta^\mu \delta_\alpha^\nu\right)$,$ T_{\alpha \beta}^{\mu \nu} \equiv \frac{1}{4} \eta^{\mu \nu} \eta_{\alpha \beta}$, $ C_{\alpha \beta}^{\mu \nu} \equiv \frac{1}{2} \left( \eta^{\mu \nu} \frac{q_\alpha q_\beta}{q^2} + \frac{q^\mu q^\nu}{q^2} \eta_{\alpha \beta} \right)$, $ J_{\alpha \beta}^{\mu \nu} \equiv I_{\rho \kappa}^{\mu \nu} \frac{q_\sigma q^\rho}{q^2} I_{\alpha \beta}^{\sigma \kappa}$, and $K_{\alpha \beta}^{\mu \nu} \equiv \frac{q^\mu q^\nu}{q^2} \frac{q_\alpha q_\beta}{q^2}$. The propagator \eqref{8} represents the most general form consistent with unitarity. It only propagates the standard massless transverse spin-2 graviton, without introducing any additional d.o.f. This is due to the fact that exponents of entire functions are themselves special entire functions that do not have zeros, thus preventing the emergence of new poles and, consequently, any new physical d.o.f. Additionally, it is crucial to note that for the propagator to remain well-behaved, the correct forms of only two factors, corresponding to the $I_{\m\n\al\bet}$ and $T_{\m\n\al\bet}$ components, are required. This explains why one function out of three $F_4(\Box)$ can be put to zero from the point of view of unitarity. Furthermore, unitarity is explicitly preserved, as the optical theorem is trivially satisfied at the tree level, namely
\be
\label{9}
2 \text{Im}\{T(q)^{\m\n}O^{-1}_{\m\n\al\bet}T(q)^{\al\bet}\}=2\pi \text{Res}\{T(q)^{\m\n}O^{-1}_{\m\n\al\bet}T(q)^{\al\bet}\}|_{q^2=0}>0,
\ee
the $T^{\mu\nu}$ represents the energy-momentum tensor of matter field \cite{Koshelev:2017ebj}. In particular, due to the conservation condition, it is sufficient to consider only the $I_{\m\n\al\bet}$ and $T_{\m\n\al\bet}$ components of the propagator.
\subsection{ Power counting in Minkowski Spacetime }
We now revisit the power counting analysis of quantum divergences as presented in Refs.\cite{Kuzmin:1989sp,Tomboulis:1997gg,Tomboulis:2015esa,Eran:1998pga,Modesto:2015lna}. In the UV regime, the propagator in momentum space exhibits the following scaling behavior
\be
\label{10}
O^{-1}\sim\frac{1}{q^{2\gamma+D}}.
\ee

The vertices can be organized into different sets, some of which may involve the entire functions $e^{ H_\ell(z)}$, while others may not. However, to establish a bound on the quantum divergences, it is sufficient to focus on the leading operators in the UV regime. These operators scale inversely with the propagator, thereby yielding upper bounds on the superficial degree of divergence for any graph $G$ \cite{Kuzmin:1989sp,Tomboulis:2015esa,Modesto:2011kw,Modesto:2014lga}.
\be
\label{11}
\omega(G)=D L+(V-I)(2\gamma+D)
\ee
in a spacetime of even or odd dimensionality $D$. Here, $V$ represents the number of vertices, $I$ denotes the number of internal lines, and $L$ stands for the number of loops. Furthermore, using the topological relation $L-1=I-V$, we can simplify the expression as
\be
\label{12}
\omega(G)=D-2\gamma (L-1).
\ee

Thus, if $ \gamma > \frac{D}{2}$ (corresponding to Eq.\eqref{3}), only one-loop divergences remain in the theory, implying that higher-loop divergences are effectively suppressed. As a result, the theory is super-renormalizable \cite{Kuzmin:1989sp,Tomboulis:2015esa,Modesto:2011kw,Modesto:2014lga,Modesto:2015lna}. In this case, only a finite number of operators, each with a mass dimension up to $D$, need to be included in the action to guarantee renormalizability in even-dimensional spacetime. This ensures that the theory remains well-defined at high energies, with a controlled and finite set of divergences, thus avoiding the issues associated with non-renormalizable theories.

\section{The Metric corrections from Tree-Level Scattering Amplitudes}
It is an intriguing concept that the Schwarzschild-Tangherlini metric can be derived from scattering amplitudes and Feynman diagrams \cite{Bjerrum-Bohr:2018xdl,Jakobsen:2020ksu,Mougiakakos:2020laz}. 
In this section, we examine the connection between super-renormalizable gravity and the vertex function of massive scalar particles interacting with gravitons. Specifically, we investigate whether it is possible to derive the gravitational metric from the scattering amplitudes. Besides, it is worth mentioning that our primary focus is on the classical limit of QG in flat spacetime, where, within this limit, the effects of ghost fields and divergent terms can be neglected. The classical limit of QFT is typically defined as the regime where $\hbar \to 0$. This is traditionally understood as the classical variational principle, $\delta S = 0$. However, even in this limit, Feynman diagrams at any loop order can still make contributions. A detailed analysis of the classical limit of QFT can be found in Ref.\cite{Kosower:2018adc}, with specific attention to gravity in Refs.\cite{Bjerrum-Bohr:2018xdl,Cheung:2020gyp,Bjerrum-Bohr:2019kec}. Furthermore, we consider two types of particles: massive scalar particles and gravitons. The massive scalars are treated as point particles, with their momenta remaining finite, while gravitons behave like waves in the classical theory, with their momenta approaching zero. These conclusions hold for both external and internal loop momenta. In this sense, the classical limit is analogous to a long-range limit.

The vertex rules come from the expansion of the matter term of the action which is
\be
\label{13}
S_{\text{matter}}=\frac{1}{2}\int \dif^D x \sqrt{-g}\(\na_{\m}\phi\na^{\m}\phi-m^2\phi^2   \),
\ee
where the scalar fields are treated as neutral, non-rotating point particles. To simplify, we represent the matter and gravitational field using minimal coupling. Then, we proceed by expanding the action in terms of powers of the fields
\be
\label{14}
\begin{aligned}
S_{\text{matter}}& \approx  \frac{1}{2} \int d^D x\left(\eta^{\mu \nu} \phi_{, \mu} \phi_{, \nu}-m^2 \phi^2\right) \\
& -\frac{1}{2} \int d^D x\left(h^{\mu \nu} \phi_{, \mu} \phi_{, \nu}-\frac{1}{2} h_\mu^\mu\left(\phi^{, \nu} \phi_{, \nu}-m^2 \phi^2\right)\right).
\end{aligned}
\ee
From the $\phi^2h$ action we get tree-lever vertex rule proportional to $-i \kappa V^{\m\n}_{\text{vertex}}$, where
\be
\label{15}
V_{\text{vertex}}^{\m \n}(k, k-q, m)\equiv I_{\alpha \beta}^{\mu \nu} k^\alpha (k-q)^\beta-\frac{k(k-q)-m^2}{2} \eta^{\mu \nu}
\ee
with tree-level Feynman diagram
\begin{figure}[H]
\centering
\begin{minipage}{0.8\textwidth}
\centering
\includegraphics[scale=0.08,angle=0]{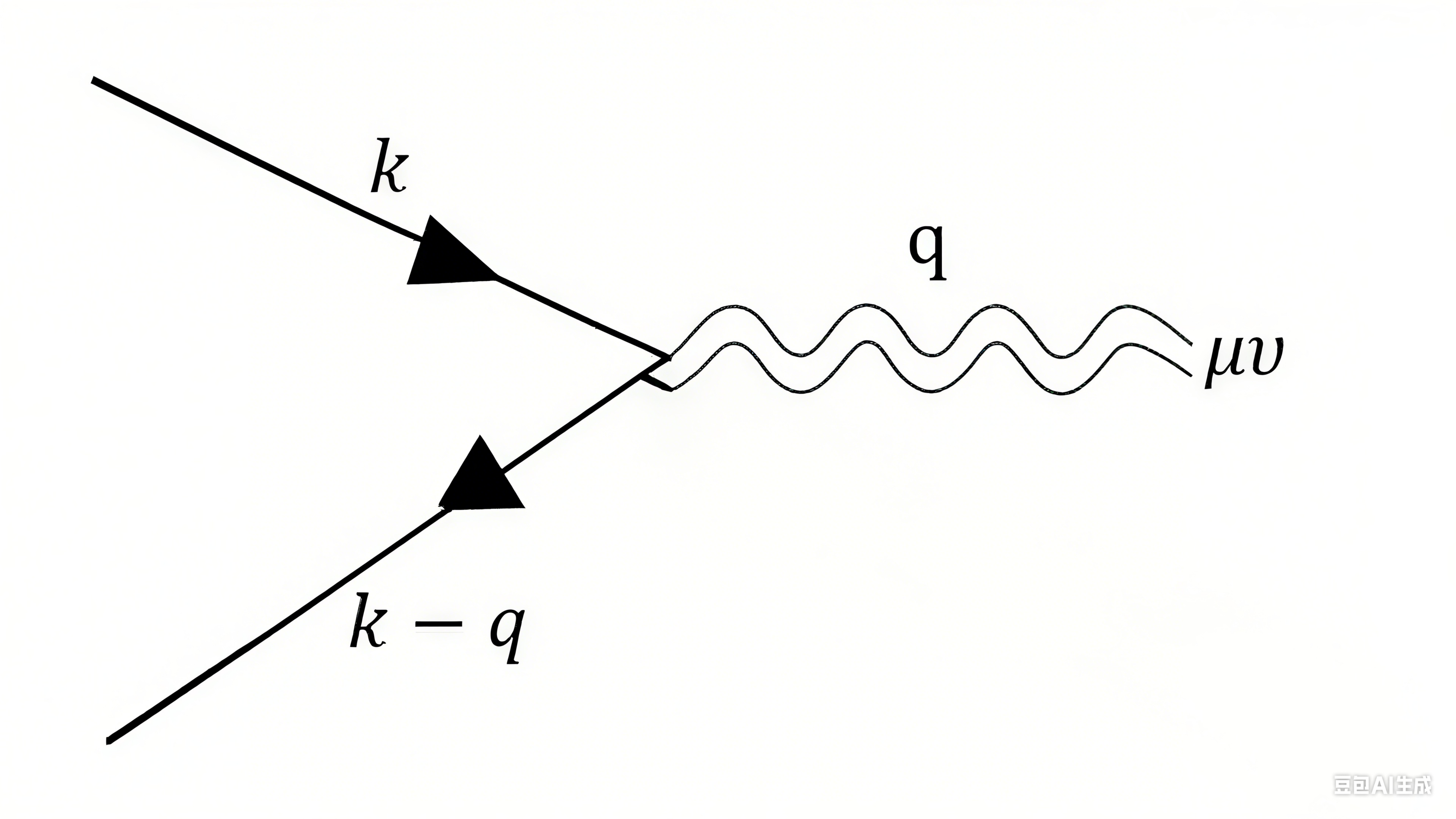}
\end{minipage}
\caption{\label{Fig.1} The tree-level contribution to the three-point vertex function involves the graviton momentum $q$, and the four-momentum $k$ of the on-shell particle, which satisfies the condition $k^2 = m^2$.} 
\end{figure}

Moreover, the Lorentz invariance of the perturbative QFT permits the use of an arbitrary inertial reference frame. To streamline the analysis, we introduce a notation that separates tensors into parallel and orthogonal components relative to the momentum $k_{\mu}$ of the scalar particle. This leads to the definition of the following projection operators
\be
\1\{\begin{split}
\label{16}
&\eta^{||}_{\m\n}=\frac{k^{\m}k^{\n}}{m^2},\\
&\eta^{\bot}_{\m\n}=n_{\m\n}-\frac{k^{\m}k^{\n}}{m^2}.
\end{split}\2.
\ee

These two operators are independent, and their sum is equal to $\eta_{\mu\nu}$. In the inertial frame of $k_{\mu}$, these components simplify significantly, becoming diagonal and representing the time and spatial components, respectively. 
As depicted in Fig. \ref{Fig.1}, the amplitude in this regime encodes the metric perturbation $h_{\mu\nu}$ generated by the point-like source and its gravitational field. This amplitude, denoted $\mathcal{M}^{\mu\nu}_{\text{vertex}}$, incorporates both the energy-momentum pseudotensor and gauge-fixing contributions. A relation derived in Ref.\cite{Jakobsen:2020ksu} links the gauge-invariant part of the scattering amplitude to the physical pseudotensor as 

\be
\label{17}
2\pi\delta\(kq\)\mathcal{M}^{\m\n}_{\text{vertex}}=-\kappa \tilde{\tau}^{\m\n}+\frac{1}{\xi}\tilde{H}^{\m\n}_{\text{non-linear}},
\ee
where $\tilde{\tau}^{\m\n}$ represents the combined contribution from the matter ($T_{\m\n}$) and the nonlinear gravitational field. To zeroth order, it corresponds to the energy-momentum tensor of  point particle in special relativity, while loop corrections account for the energy-momentum contribution from the surrounding, self-interacting gravitational field. $\tilde{H}^{\m\n}_{\text{non-linear}}$ denotes the total contribution from the nonlinear gauge-fixing term. However, we adopt Eq.\eqref{6} as the gauge-fixing term, which does not contribute any nonlinear terms. Therefore, this term can be safely neglected in our calculations. The linear term appears as a dynamical propagator, which contributes to Eq.\eqref{8}. Furthermore, due to the conservation of momentum in both the initial and final states, the constraint $\delta(kq) = \delta(mq_{||})$ must be imposed to properly define the integral in classical limit. To obtain the correction to the metric, we perform a Fourier transform of the above expression and combine it with the equations of motion, resulting in
\be
\label{18}
g_{\mu \nu}=\eta_{\mu \nu}-\frac{\kappa}{2} \int \frac{\dif^D q \delta(k q) e^{-i q x}}{(2 \pi)^{D-1}} O^{-1}_{\al\bet\m\n} \mathcal{M}_{\mathrm{vertex}}^{\alpha \beta} .
\ee

For GR, it takes the form $O^{-1}_{\alpha\beta\mu\nu} = \frac{I_{\alpha\beta\mu\nu} - \frac{1}{D-2} \eta_{\mu\nu} \eta_{\alpha\beta}}{q^2 + i\epsilon}$, with the corresponding Schwarzschild-Tangherlini  solution already derived in Ref.\cite{Khalaf:2023ozy}. Nevertheless, we focus on super-renormalizable gravity, where the action is given by Eq.\eqref{1}. For simplicity, we examine a particular class of form factors, which can be chosen as
\be
\1\{\begin{split}
\label{19}
&H_0(\Box)=H_2(\Box)=-\frac{\Box}{M^2_{\ast}}\Rightarrow F_1(\Box)=-\frac{F_2(\Box)}{2}=\frac{e^{-\frac{\Box}{M^2_{\ast}}}-1 }{\Box} ,\\
&\omega(\Box)=e^{-\frac{\Box}{M^2_{\ast}}}.
\end{split}\2.
\ee

It is straightforward to show that this choice of form factor simplifies the propagator Eq.\eqref{8} to the following form
\be
\label{20}
\begin{split}
O^{-1}_{\m\n\al\bet}&=\frac{e^{-\frac{q^2}{M^2_{\ast}}}}{q^2} I_{\m\n\al\bet}- \frac{4e^{-\frac{q^2}{M^2_{\ast}}}}{(D-2)q^2}T_{\m\n\al\bet}+\frac{2e^{-\frac{q^2}{M^2_{\ast}}}}{(D-2)q^2}C_{\m\n\al\bet}\\
&+ \frac{(2\xi-2)e^{-\frac{q^2}{M^2_{\ast}}}}{q^2}J_{\m\n\al\bet}  +\frac{e^{-\frac{q^2}{M^2_{\ast}}}}{q^2}\( \frac{D-2}{D-1}-\frac{1}{(D-2)(D-1)}-\frac{3\xi}{2}\)  K_{\m\n\al\bet}.
\end{split}
\ee

Subsequently, based on the Feynman rules in Eq.\eqref{15}, it can be demonstrated that the tree-level scattering amplitude in classical limit is given by $\mathcal{M}^{\m\n}_{\text{tree}}=-\kappa m^2\eta^{||}_{\m\n}$. Based on the coupling between the propagator and the scattering amplitudes \eqref{18}, we calculate the first-order correction to the metric tensor as
\be
\1\{\begin{split}
\label{21}
&h^{(G_{N})}_{\al\bet}=  T_1\eta^{||}_{\al\bet}+T_2 \eta^{\bot}_{\al\bet},\\
&T_1\equiv  \frac{m\kappa^2}{2}\int \frac{\dif^{D-1} q_{\bot} }{(2\pi)^{D-1}}\(  \frac{(D-3)e^{-\frac{q^2_{\bot}}{M^2_{\ast}}}}{(D-2)q^2_{\bot}}       \)e^{-iq_{\bot}x_{\bot}}   ,\\
&T_2\equiv-\frac{m\kappa^2}{2}\int \frac{\dif^{D-1} q_{\bot} }{(2\pi)^{D-1}}\(   \frac{e^{-\frac{q^2_{\bot}}{M^2_{\ast}}}}{(D-2)q^2_{\bot}}       \)e^{-iq_{\bot}x_{\bot}},
\end{split}\2.
\ee
where we have used the $\delta$ function to express $q^{\m}$ and $x^\m$ in terms of $q^{\m}_{\bot}$ and $x^{\m}_{\bot}$ (Since $q^{\m}=q^{\m}_{||}+q^{\m}_{\bot}, x^{\m}=x^{\m}_{||}+x^{\m}_{\bot}$). Observe that the integrand includes an exponential function, and we apply the following formula
\be
\label{22}
\begin{split}
\int \frac{\dif^{D-1} q_{\bot} }{(2\pi)^{D-1}}\(  \frac{e^{-\frac{q^2_{\bot}}{M^2_{\ast}}}}{q^2_{\bot}}       \)e^{-iq_{\bot}x_{\bot}}= -\frac{1}{4\pi^{\frac{D-1}{2}}r^{D-3}}\[\Gamma\(\frac{D-3}{2}\)-\Gamma\(\frac{D-3}{2},\frac{1}{4}M^2_{\ast}r^2 \)\],
\end{split}
\ee
where $\Gamma\left(\frac{D-3}{2}, \frac{1}{4}M^2_{\ast}r^2\right)$ represents the upper incomplete gamma function of order $ \frac{D-3}{2}$, and the variable $r$, defined as $r^2 = -x^2_{\bot}$ in the reference frame of $k_{\mu}$. By substituting Eq.\eqref{22} into Eq.\eqref{21}, we ultimately obtain
\be
\label{23}
\begin{split}
h^{(G_{N})}_{\al\bet}&=-\frac{(D-3)4\pi G_N m}{(D-2)\pi^{\frac{D-1}{2}}r^{D-3}}\[\Gamma\(\frac{D-3}{2}\)-\Gamma\(\frac{D-3}{2},\frac{1}{4}M^2_{\ast}r^2 \)\]\eta^{||}_{\al\bet}\\
&+ \frac{4\pi G_N m}{(D-2)\pi^{\frac{D-1}{2}}r^{D-3}}\[\Gamma\(\frac{D-3}{2}\)-\Gamma\(\frac{D-3}{2},\frac{1}{4}M^2_{\ast}r^2 \)\]\eta^{\bot}_{\al\bet},
\end{split}
\ee
where we derived the first-order correction to the metric in terms of the Newtonian constant $G_N$. Based on Newtonian potential $\Phi\equiv\frac{h_{00}}{2}$, the $D$ dimensional  $\mathcal O(G_N)$ order potential function can be written as
\be
\label{24}
\begin{split}
\Phi^{(G_N)}(r)=-\frac{(D-3)2\pi G_N m}{(D-2)\pi^{\frac{D-1}{2}}r^{D-3}}\[\Gamma\(\frac{D-3}{2}\)-\Gamma\(\frac{D-3}{2},\frac{1}{4}M^2_{\ast}r^2 \)\].
\end{split}
\ee

In particular, in the case of $D=4$, the potential can be derived as
\be
\label{25}
\begin{split}
\Phi^{(G_N)}(r)=-\frac{G_N m}{r}\text{erf}\(\frac{M_{\ast} r}{2}\),
\end{split}
\ee
where the error function, denoted as $\text{erf}(x)$, is defined by $\text{erf}(x) = \frac{2}{\sqrt{\pi}} \int_0^x e^{-t^2} \, dt$. This result aligns with earlier work in \cite{Biswas:2011ar,Burzilla:2020utr} obtained with linear perturbation. The error function encodes the nonlocal smearing of the gravitational interaction: at large distances $\lim_{x \to \infty} \text{erf}(x) = 1$, the potential reduces to the standard Newtonian form $-\frac{G_N m}{r}$, recovering GR in the infrared region. 

\section{The Metric corrections from One-Loop Scattering Amplitudes}
In this section, we will calculate the second-order correction to the metric in the classical limit, stemming from the one-loop scattering amplitude. The $(G_N)^2$ contribution to the metric is generated by the triangle one-loop diagram depicted in
Fig.2. 
Other one-loop topologies, devoid of such non-analyticity, do not contribute to the classical potential, thereby establishing the triangle diagram as the sole contributor to the $\mathcal{O}(G_N^2)$ singularity resolution \cite{Bjerrum-Bohr:2018xdl}.
\begin{figure}[H]
\centering
\begin{minipage}{0.8\textwidth}
\centering
\includegraphics[scale=0.5,angle=0]{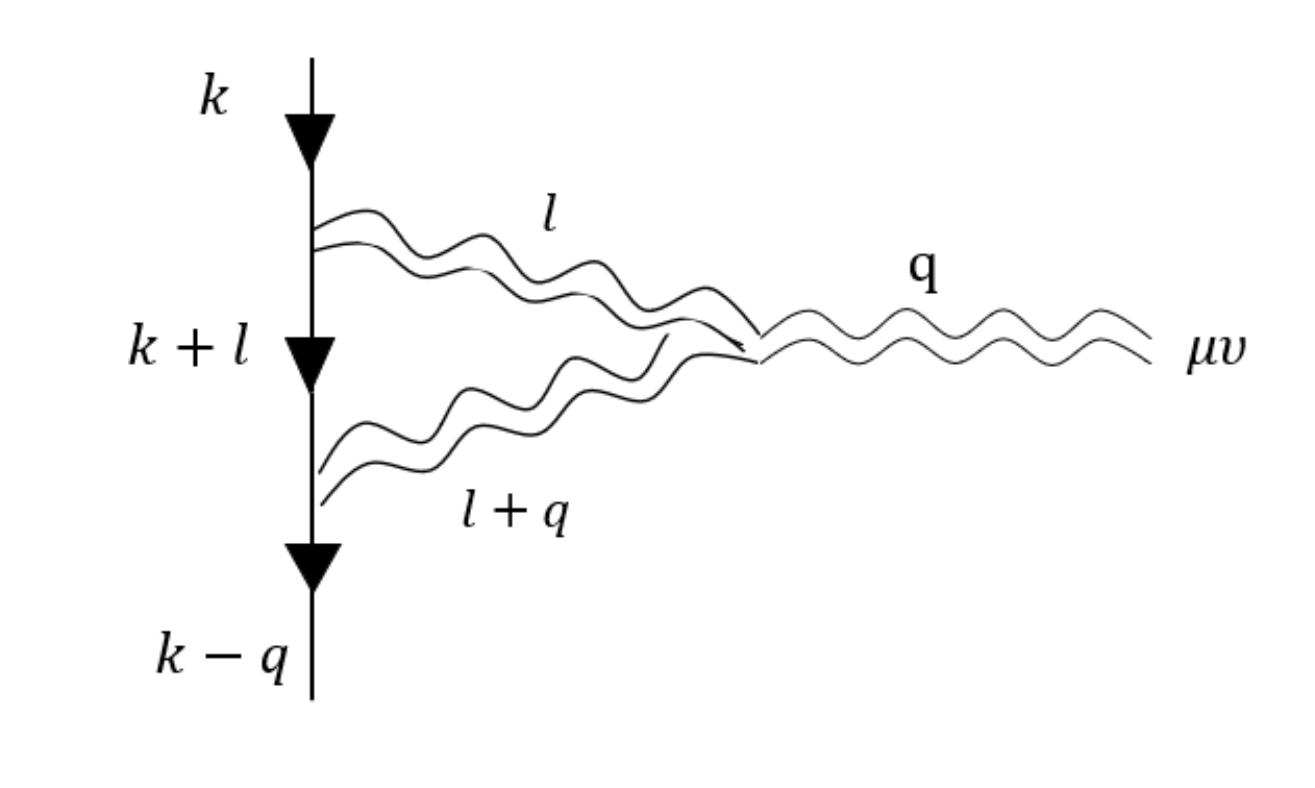}
\end{minipage}
\caption{\label{Fig.2} The one-loop contribution to the three-point vertex function involves $q$, and $k$. The loop momentum is denoted by $l$.}
\end{figure}
Employing the Feynman rules in Appendix A for the three-graviton vertex and for the interaction between two scalar particles and one graviton, together with the propagator from Eq. \eqref{20} and the metric correction in Eq. \eqref{18}, we obtain
\be
\label{26}
\begin{split}
i\mathcal{M}^{\m\n}_{\text{vertex}}&\equiv i\mathcal{M}^{\m\n}_{\text{GR}}+i\mathcal{M}^{\m\n}_{\text{Nonlocal}}\\
&=-2m^4\kappa^3\int\frac{{\dif^D l}}{(2\pi)^D}\frac{e^{-\frac{l^2}{M^2_{\ast}}}e^{-\frac{\(l+q_{\bot}\)^2}{M^2_{\ast}}}f_{\al\bet}f_{\gamma\delta}\(V^{\al\bet\gamma\delta\m\n}_{h^3 \text{GR}} +V^{\al\bet\gamma\delta\m\n}_{h^3 \text{Nonocal}} \) }{l^2(l+q_{\bot})^2\((l+k)^2-m^2+i\epsilon\)},
\end{split}
\ee
where $V^{\al\bet\gamma\delta\m\n}_{h^3\text{GR}}$ and $V^{\al\bet\gamma\delta\m\n}_{h^3\text{Nonlocal}}$ represent the interaction terms involving the three-graviton vertices from GR and nonlocal terms, respectively. In addition, we introduce the tensor $f_{\alpha \beta}\equiv\frac{D-3}{D-2}\eta^{||}_{\al\bet}-\frac{1}{D-2}\eta^{\bot}_{\al\bet}$, which represents the scalar-graviton vertices contracted with the graviton propagator. These tensors encapsulate the tensor structure of the $\phi^2 h$ vertices after they are contracted with the graviton propagator. We employ two important simplifications, which arise within the classical limit. Firstly, we neglect the factors of $l$ and $q$ in the scalar-graviton vertex. Secondly, we omit the momentum-dependent part of the graviton propagator. The first simplification is analogous to the treatment of tree-level diagram, where graviton momenta are considered negligible relative to the scalar momenta. The second simplification, involving the momentum dependence of the propagator, is less trivial, as graviton momenta appear in both the numerator and denominator of the propagator expression. This can be confirmed by examining the corresponding integrals, though it is also clear from an intuitive perspective. Since scalar-graviton vertices represent energy-momentum and are conserved, they vanish when contracted with the momentum from the propagator.


Furthermore, focusing on the tensor part of the integral involving the three-graviton vertex from the GR contribution $\(V^{\alpha\beta\gamma\delta\mu\nu}_{h^3 \text{GR}}\)$ and factors of $f_{\alpha\beta}$,  we express the amplitude from the GR part as
\be
\label{28}
\begin{split}
i\mathcal{M}^{\m\n}_{\text{GR}}=-2m^4\kappa^3f_{\al\bet}f_{\gamma\delta}\(U^{\m\n\al\bet\rho\gamma\delta\sigma}_{\text{GR}}\(I^{\text{GR}}_{\rho\sigma}+I^{\text{GR}}_{\rho}q_{\bot\sigma}\)+U^{\al\bet\gamma\delta\rho\m\n\sigma}_{\text{GR}}q_{\bot\rho}q_{\bot\sigma}I^{\text{GR}}\),
\end{split}
\ee
where $U$-tensor is defined in Appendix A, and the quantities $I^{\text{GR}}_{\rho\sigma}$, $I^{\text{GR}}_{\rho}$, and $I^{\text{GR}}$ are given by
\be
\label{29}
\begin{split}
I^{\text{GR}}_{a_1...a_n}\equiv\int\frac{{\dif^D l}}{(2\pi)^D}\frac{e^{-\frac{l^2}{M^2_{\ast}}}e^{-\frac{\(l+q_{\bot}\)^2}{M^2_{\ast}}} l_{a_1}...l_{a_n}}{l^2(l+q_{\bot})^2\((l+k)^2-m^2+i\epsilon\)}.
\end{split}
\ee

Using the integral formulas provided in Appendix B, we can show that the term $I^{\text{GR}}_{\rho\sigma}+I^{\text{GR}}_{\rho}q_{\bot\sigma}$ in Eq.\eqref{28} can be expressed as
\be
\label{30}
\begin{split}
I^{\text{GR}}_{\rho\sigma}+I^{\text{GR}}_{\rho}q_{\bot\sigma}=\frac{iq^2_{\bot}N_{D-1}}{16\(D-2\)m}\(\(D-3\)\frac{q_{\bot\rho}q_{\bot\sigma}}{q^2_{\bot}}+\eta^{\bot}_{\rho\sigma}\)+\frac{i N_D}{4m^2}\(k_{\rho}q_{\bot\sigma}-q_{\bot\rho}k_{\sigma}\), \end{split}
\ee
here the integrals for $N_{D-1}$ and $N_{D}$ are also defined in Appendix B. By applying the results above and the vertex rule, we obtain the final contribution from the GR part as
\be
\label{31}
\begin{split}
i\mathcal{M}^{\m\n}_{\text{GR}}=\frac{i m^3\kappa^3q^2_{\bot}N_{D-1}}{2}\( \frac{7D-25}{8(D-2)}\eta_{||}^{\m\n}-\frac{5D^2-18D+17}{8(D-2)}\(\eta^{\m\n}_{\bot}- \frac{q_{\bot\rho}q_{\bot\sigma}}{q^2_{\bot}}  \)\).
\end{split}
\ee

Subsequently, using the vertex rule for nonlocal terms provided in Appendix A (denoted as $V^{\al\bet\gamma\delta\m\n}_{h^3\text{Nonlocal}}$), we obtain
\be
\label{32}
\begin{split}
i\mathcal{M}^{\m\n}_{\text{Nonlocal}}=&-2m^4\kappa^3f_{a_1a_2}f_{b_1b_2}\[U^{\m\n a_1a_2abb_1b_2cd}_{1\text{Nonlocal}}\(I^{(1)}_{abcd}+I^{(1)}_{abc}q_{\bot d}+I^{(1)}_{abd}q_{\bot c}+I^{(1)}_{ab}q_{\bot c}q_{\bot d} \)\right.\\
&+\left.  U^{ a_1a_2b_1b_2ab\m\n cd}_{1\text{Nonlocal}}\(I^{(2)}_{ab}q_{\bot c}q_{\bot d}+I^{(2)}_{a}q_{\bot b}q_{\bot c}q_{\bot d}+I^{(2)}_{b}q_{\bot a}q_{\bot c}q_{\bot d}+I^{(2)}q_{\bot a}q_{\bot b}q_{\bot c}q_{\bot d} \) \right.\\
&\left.+U^{ b_1b_2\m\n ab a_1a_2 cd}_{1\text{Nonlocal}}I^{(3)}_{cd}q_{\bot a}q_{\bot b}-U^{a_1a_2ab_1b_2b\m\n cd}_{2\text{Nonlocal}}\(I^{(2)}_{ab}q_{\bot c}q_{\bot d}+I^{(2)}_{a}q_{\bot b}q_{\bot c}q_{\bot d} \)\right.\\
&\left.+U^{\m\n aa_1a_2bb_1b_2cd}_{2\text{Nonlocal}}\(I^{(1)}_{bcd}q_{\bot a}+I^{(1)}_{bc}q_{\bot a}q_{\bot d}+I^{(1)}_{bd}q_{\bot a}q_{\bot c}+I^{(1)}_{b} q_{\bot a}q_{\bot c}q_{\bot d}  \)\right.\\
&\left.- U^{b_1b_2a\m\n ba_1a_2cd}_{2\text{Nonlocal}}\(I^{(3)}_{acd}q_{\bot b}+q_{\bot a}q_{\bot b}I^{(3)}_{cd} \) \],
\end{split}
\ee
where we define three types of integrals, which primarily differ due to the contributions from the form factor $F_1(\Box)$. These integrals can be written as
\be
\1\{\begin{split}
\label{33}
&I^{(1)}_{a_1...a_n}\equiv-\int\frac{{\dif^D l}}{(2\pi)^D}\frac{e^{-\frac{l^2}{M^2_{\ast}}}e^{-\frac{\(l+q_{\bot}\)^2}{M^2_{\ast}}}\(e^{\frac{\(l+q_{\bot}\)^2}{M^2_{\ast}}} -1\) l_{a_1}...l_{a_n} }{l^2(l+q_{\bot})^4\((l+k)^2-m^2+i\epsilon\)},\\
&I^{(2)}_{a_1...a_n}\equiv-\int\frac{{\dif^D l}}{(2\pi)^D}\frac{e^{-\frac{l^2}{M^2_{\ast}}}e^{-\frac{\(l+q_{\bot}\)^2}{M^2_{\ast}}}\(e^{\frac{q^2_{\bot}}{M^2_{\ast}}} -1\) l_{a_1}...l_{a_n} }{l^2q^2_{\bot}(l+q_{\bot})^2\((l+k)^2-m^2+i\epsilon\)},\\
&I^{(3)}_{a_1...a_n}\equiv-\int\frac{{\dif^D l}}{(2\pi)^D}\frac{e^{-\frac{l^2}{M^2_{\ast}}}e^{-\frac{\(l+q_{\bot}\)^2}{M^2_{\ast}}}\(e^{\frac{l^2}{M^2_{\ast}}} -1\) l_{a_1}...l_{a_n} }{l^4(l+q_{\bot})^2\((l+k)^2-m^2+i\epsilon\)}.\\
\end{split}\2.
\ee
Similarly, based on the definition of the full amplitude contribution from nonlocal terms, we obtain
\be
\1\{\begin{split}
\label{34}
&i\mathcal{M}^{\m\n}_{\text{Nonlocal}}=-\frac{i m^2\kappa^3q^2_{\bot}\(k^{\m}q^{\n}_{\bot}+k^{\n}q^{\m}_{\bot}\)}{16(D-2)^2(D-1)}T_3+\frac{i m^3\kappa^3q^4_{\bot}}{16(D-2)^3}\(T_4\eta^{\m\n}_{||}+T_5\eta^{\m\n}_{\bot}+T_6\frac{q^{\m}_{\bot}q^{\n}_{\bot}}{q^2_{\bot}} \),\\
&T_3\equiv\(48(D-1)-9D^2\)N^{(1)}_{D}+  8(D-1)^2N^{(2)}_D+\(8D^3-43D^2+72D-32\)N^{(3)}_{D},\\
&T_4\equiv -4\(D^2-5D+6\)N^{(1)}_{D-1}+\(20D^3-140D^2+342D-284 \)N^{(2)}_{D-1}-4\(D^2-6D+8 \)N^{(3)}_{D-1} ,\\
&T_5\equiv \(4D-11\)N^{(1)}_{D-1}+\(20D^3-148D^2+382D-334 \)N^{(2)}_{D-1}+\(8D-19\)N^{(3)}_{D-1} ,\\
&T_6\equiv3\(D-1\)N^{(1)}_{D-1}-\(12D^3-92D^2+242D-214  \)N^{(2)}_{D-1}+\(13-5D\)N^{(3)}_{D-1} ,\\
\end{split}\2.
\ee
the integrals $N^{(i)}_{D},N^{(i)}_{D-1},i=1,2,3$ are given by Appendix B. From the above expression, it is evident that the contribution from the nonlocal terms introduces additional term involving $k^{\mu} q^{\nu}_{\bot} + k^{\nu} q^{\mu}_{\bot}$ compared to the GR. This indicates that such term contribute to correction in the spacetime cross terms of the metric. Furthermore, we find that the nonlocal terms also include an extra $q^2_{\bot}$ term in the numerator compared to the GR term. This suggests that, in the classical limit, the GR contribution will dominate. It also implies that the theory has a certain range of feasible scales within which it is valid.

Ultimately, by substituting Eq.\eqref{31} and Eq.\eqref{34} into Eq.\eqref{18}, we obtain the corrections to the metric and Newtonian potential up to second order in $(G_{N})^2$ within the one-loop approximation. These corrections can be expressed  as
\be
\1\{\begin{split}
\label{35}
&h^{(G_N)^2}_{\m\n}\equiv h^{(G_N)^2\text{GR}}_{\m\n}+h^{(G_N)^2\text{Nonlocal}}_{\m\n},\\
& h^{(G_N)^2\text{GR}}_{\m\n}=-\frac{\kappa^4m^2}{2}\int\frac{\dif^{D-1}q_{\bot}}{(2\pi)^{D-1}}e^{-i q_{\bot}x_{\bot}}e^{-\frac{q^2_{\bot}}{M^2_{\ast}}}N_{D-1}\[ \frac{183D+5D^3-61D^2-151}{16(D-2)^2}\eta^{||}_{\m\n}\right.\\
&\left.+\frac{25-7D}{16(D-2)^2}\eta^{\bot}_{\m\n}+\frac{(17+D)(5D-18)}{16(D-2)}\frac{q_{\m}q_{\n}}{q^2_{\bot}}          \],\\
&h^{(G_N)^2\text{Nonlocal}}_{\m\n}=-\frac{\kappa^4m^2}{2}\int\frac{\dif^{D-1}q_{\bot}}{(2\pi)^{D-1}}e^{-i q_{\bot}x_{\bot}}e^{-\frac{q^2_{\bot}}{M^2_{\ast}}}\[\frac{3N^{(1)}_{D}\(16-16D+3D^2\)}{16m(D-1)(D-2)^2}   \(k_{\m}q_{\bot\n}+k_{\n}q_{\bot\m}\)          \right.\\
&\left.-\frac{N^{(2)}_{D}(D-1)}{2m (D-2)^2}            \(k_{\m}q_{\bot\n}+k_{\n}q_{\bot\m}\)   -\frac{N^{(3)}_{D}\(8 D^3-43D^2+72D-32\)}{16m(D-1) (D-2)^2}            \(k_{\m}q_{\bot\n}+k_{\n}q_{\bot\m}\) \right.\\
&\left.-q^2_{\bot}N^{{(1)}}_{D-1}\( \frac{D^2-5D+8}{4(D-2)^3}\eta^{||}_{\m\n}-\frac{4D-19}{16(D-2)^3}\eta^{\bot}_{\m\n}  -\frac{3(D-1)}{16(D-2)^3}\frac{q_{\bot\m}q_{\bot\n}}{q^2_{\bot}}    \)\right.\\
&\left.-q^2_{\bot}N^{{(3)}}_{D-1}\( \frac{D^2-5D+8}{4(D-2)^3}\eta^{||}_{\m\n}-\frac{4D-19}{16(D-2)^3}\eta^{\bot}_{\m\n}  -\frac{13-5D}{16(D-2)^3}\frac{q_{\bot\m}q_{\bot\n}}{q^2_{\bot}}    \)\right.\\
&\left. -q^2_{\bot}N^{{(2)}}_{D-1}\( \frac{76+13D+10D^2}{8(D-2)^3}\eta^{||}_{\m\n}-\frac{7D-14D^2-101}{8(D-2)^3}\eta^{\bot}_{\m\n}  -\frac{107-121D+46D^2-6D^3}{8(D-2)^3}\frac{q_{\bot\m}q_{\bot\n}}{q^2_{\bot}}    \)      \],
\end{split}\2.
\ee
with $(G_N)^2$ order potential function 
\be
\label{36}
\begin{split}
\Phi^{(G_N)^2}(r)&=-\frac{\kappa^4m^2}{4}\int\frac{\dif^{D-1}q_{\bot}}{(2\pi)^{D-1}}e^{-i q_{\bot}x_{\bot}}e^{-\frac{q^2_{\bot}}{M^2_{\ast}}}\[N_{D-1}\frac{183D+5D^3-61D^2-151}{16(D-2)^2}\right.\\
&\left.-q^2_{\bot}\(N^{(1)}_{D-1}+N^{(3)}_{D-1}\)\(\frac{D^2-5D+8}{4(D-2)^3}\)-q^2_{\bot}N^{(2)}_{D-1}\( \frac{76+13D+10D^2}{8(D-2)^3}\)       \]  .
\end{split}
\ee

\begin{figure}[H]
\begin{minipage}{0.95\textwidth}
\centering
\includegraphics[scale=1.2,angle=0]{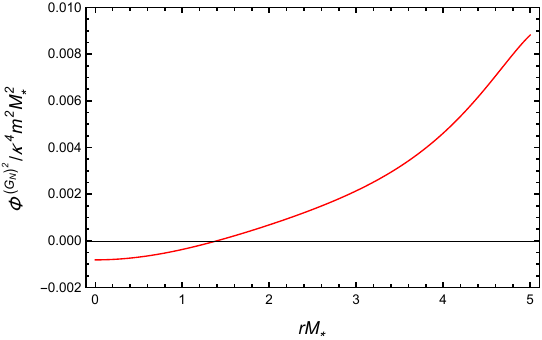}
\end{minipage}
\caption{\label{Fig.3}The Newtonian potential in $D = 4$ dimension as a function of distance.
}
\end{figure}

Based on the above equation, we find that the Newtonian potential at order $\mathcal{O}(G_N^2)$, given in Eq. (\ref{36}), is structurally governed by the spacetime dimension $D$ and four loop integrals $N_{D-1}, N^{(i)}_{D-1}$ (with $ i = 1,2,3 $), whose explicit forms are provided in Appendix B. These integrals do not admit a simple analytic expression, motivating a numerical treatment in the physically relevant case $D = 4$. As displayed in Fig. \ref{Fig.3}, the computed potential is plotted against the dimensionless radial coordinate, the one-loop correction grows smoothly with distance, matching expected long-range behavior. Most notably, the $\mathcal{O}(G^2)$ contribution to the potential varies at the level of $10^{-3}$ and shows a gradual increase with distance. At this order, the potential changes sign from negative to positive, indicating a transition from an attractive to a repulsive contribution, while the tree-level $\mathcal{O}(G)$ potential remains attractive and continues to dominate the interaction.

\section{Summary and conclusion}
In this work, we investigate the modifications to the metric corrections in $D$-dimensional spacetime within the framework of super-renormalizable gravity, focusing on the first- and second-order corrections to the gravitational constant $G_N$ through the Feynman scattering amplitudes approach. We first introduce the super-renormalizable gravity, analyzing it from two key perspectives: (1) from the viewpoint of the propagator and unitarity, where we observe that unitarity is preserved at the tree-level. and (2) from the power-counting perspective, where we identify that the primary source of divergence arises from the one-loop diagrams, thereby further confirming the super-renormalizable.

Subsequently, we analyze the static gravitational interaction derived from relativistic scattering amplitudes, focusing on both tree-level and one-loop contributions within the classical $(\hbar\rightarrow0)$ sector. At tree level, we derive analytic expressions for the metric corrections and the associated potential function in $D$-dimensional spacetime. We find that both quantities are governed by the upper incomplete gamma function $\Gamma\left(\frac{D-3}{2}, \frac{1}{4} M_\ast^{2} r^{2}\right)$, where $\Gamma(a,x)$ denotes the upper incomplete gamma function of order $a = (D-3)/2$. In four dimensions, our results reproduce the known linearized solutions of super-renormalizable gravity $\Phi^{(G_N)}(r)=-\frac{G_N m}{r}\text{erf}\(\frac{M_{\ast} r}{2}\)$. Since our analysis focuses on the large-distance regime, we further observe that in the limit of large radial separation $r$, the potential smoothly approaches the behavior predicted by GR.

At one-loop order, we derive the $\mathcal{O}(G^2)$ corrections to the metric components and to the associated potential function in $D$-dimensional spacetime. The loop contributions involve nontrivial vertex structures, denoted by $V^{\al\bet\gamma\delta\m\n}_{h^3\text{GR}}$ and $V^{\al\bet\gamma\delta\m\n}_{h^3\text{Nonlocal}}$, whose explicit analytic expressions are presented in the Appendix A. To organize the calculation, we introduce three classes of integrals that encode the nonanalytic dependence on the momentum transfer $q$ responsible for the long-range classical behavior of the interaction. Owing to the complexity of these integrals, a analytic evaluation is not available, and we therefore perform a numerical analysis in four-dimensions. The numerical results indicate that the $\mathcal{O}(G^2)$ contribution to the potential varies at the level of $10^{-3}$ and increases gradually with the radial distance $r$. Moreover, at this order the potential exhibits a change of sign from negative to positive values, corresponding to a transition from an attractive to a repulsive contribution, while the tree-level $\mathcal{O}(G)$ potential remains attractive and continues to dominate the overall interaction.

Additionally, we emphasize that the present analysis is restricted to the classical limit of relativistic scattering amplitudes, which isolates the nonanalytic contributions responsible for the long-range behavior of the effective static gravitational
interaction. As a consequence, the ultraviolet structure of super-renormalizable gravity and the detailed behavior of the interaction at $r = 0$ are not fully addressed in this paper. These aspects are expected to be governed by analytic, contact-term contributions in momentum space, which lie beyond the scope of the amplitude-based classical sector considered here. A systematic investigation of such analytic terms and their implications for the ultraviolet behavior and potential singularity structure of the theory will be the focus of future work.

\begin{acknowledgments}
This study was supported by the National Natural Science Foundation of China (Grant No. 12333008) and the Basic Research Program of Shanxi Province (Grant No. 202503021211204).
\end{acknowledgments}

\appendix

\section{Gravitational Self-Interaction: 3-graviton vertices}

We present here the Feynman rules in the covariant harmonic gauge for the three-graviton vertex in super-renormalizable gravity, derived from Eq. \eqref{1}. We focus on the contributions from the gravitational field alone, gauge-fixing terms are omitted as they do not contribute to the three-point interaction. The vertex is conveniently written in terms of the $U_{\text{GR}}$-tensor introduced in
\be
\label{A1}
\begin{split}
S_{\text{GR}}&=\frac{2}{\kappa^2}\int \sqrt{-g}\dif^{D}x R=\kappa\int\dif^{D}x \tilde{h}_{\m\n}U^{\m\n\al\bet\rho\gamma\delta\sigma}_{\text{GR}}\p_{\rho}\tilde{h}_{\al\bet}\p_{\sigma}\tilde{h}_{\delta\gamma}\\
&=-\kappa\int\frac{\dif^Dl_{1}}{(2\pi)^D}\int\frac{\dif^Dl_{2}}{(2\pi)^D}\int\frac{\dif^Dl_{3}}{(2\pi)^D}(2\pi)^D\delta^D\(l_{1}+l_2+l_3\)\tilde{h}_{\m\n}(l_{1})U^{\m\n\al\bet\rho\gamma\delta\sigma}_{\text{GR}}\tilde{h}_{\al\bet}(l_2)l_{2\rho}l_{3\sigma}\tilde{h}_{\delta\gamma}(l_3),
\end{split}
\ee
where the tensor $U^{\mu\nu \alpha \beta \rho \gamma \delta \sigma}_{\text{GR}}$ is defined to be symmetric under the exchange of the index groups $\alpha\beta\rho \leftrightarrow \gamma\delta\sigma$, following Fourier transformation in $D$-dimensional spacetime. It further exhibits full symmetry under the individual exchanges $\mu \leftrightarrow \nu$, $\alpha \leftrightarrow \beta$, and $\gamma \leftrightarrow \delta$, and admits the explicit form
\be
\label{A2}
\begin{split}
U^{\m\n\al\bet\rho\gamma\delta\sigma}_{\text{GR}}\tilde{h}_{\m\n}\p_{\rho}\tilde{h}_{\al\bet}\p_{\sigma}\tilde{h}_{\gamma\delta}&=\frac{1}{2}\tilde{h}^{\m}_{\n}\p_{\m}\tilde{h}\p^{\n}\tilde{h}-\frac{1}{4}\tilde{h}\p_{\rho}\tilde{h}\p^{\rho}\tilde{h}+\tilde{h}^{\m}_{\n}\p_{\rho}\tilde{h}^{\n}_{\m}\p^{\rho}\tilde{h}- \tilde{h}^{\m}_{\n}\p^{\sigma}\tilde{h}^{\n}_{\m}\p_{\rho}\tilde{h}^{\rho}_{\sigma}\\
&+\frac{1}{4}\tilde{h}\p_{\rho}\tilde{h}^{\m}_{\n}\p^{\rho}\tilde{h}^{\n}_{\m}  - \tilde{h}^{\n}_{\m}\p_{\n}\tilde{h}^{\m}_{\sigma}\p^{\sigma}\tilde{h}-\tilde{h}^{\m}_{\n}\p^{\n}\tilde{h}\p_{\rho}\tilde{h}^{\rho}_{\m}+\frac{1}{2}\tilde{h} \p_{\rho}\tilde{h}^{\rho}_{\sigma}\p^{\sigma}\tilde{h}\\
&-\tilde{h}^{\m}_{\n}\p_{\sigma}\tilde{h}^{\rho}_{\m}\p^{\sigma}\tilde{h}^{\n}_{\rho}-\frac{1}{2}\tilde{h}\p_{\m}\tilde{h}^{\rho}_{\nu}\p^{\n}\tilde{h}^{\m}_{\rho}+\tilde{h}^{\m\n}\p_{\rho}\tilde{h}^{\sigma}_{\m}\p_{\sigma}\tilde{h}^{\rho}_{\n}-\frac{1}{2}\tilde{h}^{\m}_{\n}\p_{\m}\tilde{h}^{\rho}_{\sigma}\p^{\n}\tilde{h}^{\sigma}_{\rho}\\
&+2\tilde{h}^{\m}_{\n}\p^{\n}\tilde{h}^{\sigma}_{\rho}\p_{\sigma}\tilde{h}^{\rho}_{\m}        .
\end{split}
\ee

To obtain a manifestly covariant form, we symmetrize Eq. \eqref{A1} by summing over cyclic permutations of the graviton fields $\tilde{h}_{\mu\nu}(l_1)$, $\tilde{h}_{\alpha\beta}(l_2)$, and $\tilde{h}_{\gamma\delta}(l_3)$. This procedure systematically ensures the complete symmetry of the three-graviton vertex required by the underlying diffeomorphism invariance, yielding the compact expression
\be
\label{A3}
\begin{split}
S_{\text{GR}}=&-\frac{\kappa}{3}\int\frac{\dif^Dl_{1}}{(2\pi)^D}\int\frac{\dif^Dl_{2}}{(2\pi)^D}\int\frac{\dif^Dl_{3}}{(2\pi)^D}(2\pi)^D\delta^D\(l_{1}+l_2+l_3\)\(U^{\m\n\al\bet\rho\gamma\delta\sigma}_{\text{GR}}l_{2\rho}l_{3\sigma}\right.\\
&\left.+U^{\al\bet\gamma\delta\rho\m\n\sigma}_{\text{GR}}l_{3\rho}l_{1\sigma}+U^{\gamma\delta\m\n\rho\al\bet\sigma}_{\text{GR}}l_{2\rho}l_{3\sigma}\)\tilde{h}_{\m\n}(l_{1})\tilde{h}_{\al\bet}(l_2)\tilde{h}_{\delta\gamma}(l_3).
\end{split}
\ee
From the above equation, we can extract the Feynman rules for the three-graviton vertex as
\be
\label{A4}
\begin{split}
2i \kappa V^{\m\n\al\bet\gamma\delta}_{h^3\text{GR}}\(l_1,l_2,l_3\)&=-2i\kappa\(U^{\m\n\al\bet\rho\gamma\delta\sigma}_{\text{GR}}l_{2\rho}l_{3\sigma}+U^{\al\bet\gamma\delta\rho\m\n\sigma}_{\text{GR}}l_{3\rho}l_{1\sigma}+U^{\gamma\delta\m\n\rho\al\bet\sigma}_{\text{GR}}l_{2\rho}l_{3\sigma} \).
\end{split}
\ee

\begin{figure}[H]
\centering
\begin{minipage}{0.5\textwidth}
\centering
\includegraphics[scale=0.5,angle=0]{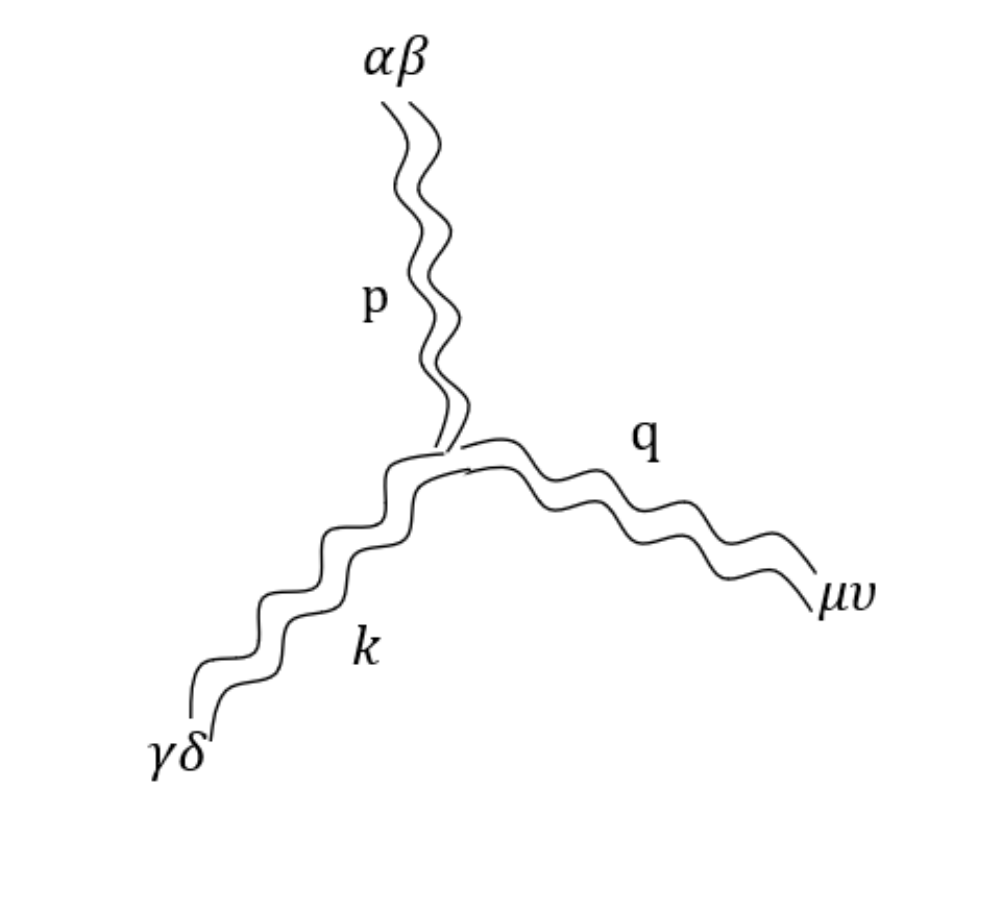}
\end{minipage}
\caption{\label{Fig. 4}
The Feynman diagram for the three-graviton vertex interaction with $2i\kappa V^{\m\n\al\bet\gamma\delta}_{h^3}\(p,q,k\)$.}
\end{figure}

Having established the GR contribution, we now turn to the vertex terms originating from the nonlocal sector of the action, which are expressed as
\be
\label{A5}
\begin{split}
S_{\text{Nonlocal}}&=\frac{2}{\kappa^2}\int \sqrt{-g}\dif^{D}x \(RF_1(\Box)R-2R_{\m\n}F_1(\Box)R^{\m\n}\)\\
&=\kappa\int\dif^Dx\( U^{\m\n a_1a_2abb_1b_2cd}_{1\text{Nonlocal}}\tilde{h}_{\m\n}\p_a\p_b\tilde{h}_{a_1a_2}F_1(\Box)\p_c\p_d\tilde{h}_{b_1b_2}\right.\\
&+\left. U^{\m\n aa_1a_2bb_1b_2cd}_{2\text{Nonlocal}}\p_a \tilde{h}_{\m\n}\p_b\tilde{h}_{a_1a_2}F_1(\Box)\p_c\p_d\tilde{h}_{b_1b_2}\),
\end{split}
\ee
with definition
\be
\label{A6}
\begin{split}
 &U^{\m\n a_1a_2abb_1b_2cd}_{1\text{Nonlocal}}\tilde{h}_{\m\n}\p_a\p_b\tilde{h}_{a_1a_2}F_1(\Box)\p_c\p_d\tilde{h}_{b_1b_2}=\tilde{h}\p_a\p_b\tilde{h}^{ab}F_1(\Box)\p_{\m}\p_{\n}\tilde{h}^{\m\n}-\tilde{h}\p_a\p_b\tilde{h}^{ab}F_1(\Box)\Box\tilde{h}\\
 &-\tilde{h}\Box\tilde{h}F_1(\Box)\p_{\m}\p_{\n}\tilde{h}^{\m\n}+\tilde{h}\Box\tilde{h}F_1(\Box)\Box\tilde{h}+4\tilde{h}_{ab}\Box\tilde{h}^{ab}F_{1}(\Box)\p_{\m}\p_{\n}\tilde{h}^{\m\n}-8\tilde{h}_{\m\n}\p^{\m}\p^{\rho}\tilde{h}^{\n}_{\rho}F_1(\Box)\p_{a}\p_{b}\tilde{h}^{ab}\\
 &+4\tilde{h}_{\m\n}\p^{\m}\p^{\n}\tilde{h}F_1(\Box)\p_{a}\p_{b}\tilde{h}^{ab}-4\tilde{h}_{\m\n}\Box\tilde{h}^{\m\n}F_1(\Box)\Box\tilde{h}+8\tilde{h}_{\m\n}\p^{\m}\p^{\rho}\tilde{h}^{\n}_{\rho}F_1(\Box)\Box\tilde{h}-4\tilde{h}_{\m\n}\p^{\m}\p^{\n}\tilde{h}F_1(\Box)\Box\tilde{h}\\
 &-\tilde{h}\p_{\m}\p_{\rho}\tilde{h}^{\rho}_{\n}F_1(\Box)\p^{\m}\p_{a}\tilde{h}^{a\n}-\tilde{h}\p_{\m}\p_{\rho}\tilde{h}^{\rho}_{\n}F_1(\Box)\p^{\n}\p_{a}\tilde{h}^{a\m}+\tilde{h}\p_{\m}\p_{\rho}\tilde{h}^{\rho}_{\n}F_1(\Box)\p^{\m}\p^{\n}\tilde{h}+\tilde{h}\p_{\m}\p_{\rho}\tilde{h}^{\rho}_{\n}F_1(\Box)\Box\tilde{h}^{\m\n}\\
 &+\tilde{h}\p_{\m}\p_{\n}\tilde{h}F_1(\Box)\p^{\m}\p_{\rho}\tilde{h}^{\rho\n}  -\frac{1}{2}\tilde{h}\p_{\m}\p_{\n}\tilde{h}F_1(\Box)\p^{\m}\p^{\n}\tilde{h}-\frac{1}{2}\tilde{h}\p_{\m}\p_{\n}\tilde{h}F_1(\Box)\Box\tilde{h}^{\m\n}+\tilde{h}\Box\tilde{h}_{\m\n}F_1(\Box)\p^{\m}\p_{\rho}\tilde{h}^{\rho\n}\\
 &-\frac{1}{2}\tilde{h}\Box\tilde{h}_{\m\n}F_1(\Box)\p^{\m}\p^{\n}\tilde{h}-\frac{1}{2}\tilde{h}\Box\tilde{h}_{\m\n}F_1(\Box)\Box\tilde{h}^{\m\n}+2\tilde{h}^{a\m}\p_a\p_{\n}\tilde{h}F_1(\Box)\Box\tilde{h}^{\n}_{\m}+2\tilde{h}^{a\m}\p_a\p_{\n}\tilde{h}F_1(\Box)\p^{\n}\p_{\m}\tilde{h}\\
 &-2\tilde{h}^{a\m}\p_a\p_{\n}\tilde{h}F_1(\Box)\p_{\m}\p_{\rho}\tilde{h}^{\rho\n}+2\tilde{h}^{a\m}\Box\tilde{h}_{a\n}F_1(\Box)\Box\tilde{h}^{\n}_{\m}+2\tilde{h}^{a\m}\Box\tilde{h}_{a\n}F_1(\Box)\p^{\n}\p_{\m}\tilde{h}-2\tilde{h}^{a\m}\Box\tilde{h}_{a\n}F_1(\Box)\p_{\m}\p_{\rho}\tilde{h}^{\rho\n}\\
&-2\tilde{h}^{a\m}\p_{\n}\p_{\rho}\tilde{h}^{\rho}_{a}F_1(\Box)\Box\tilde{h}^{\n}_{\m}-2\tilde{h}^{a\m}\p_{\n}\p_{\rho}\tilde{h}^{\rho}_{a}F_1(\Box)\p^{\n}\p_{\m}\tilde{h}+2\tilde{h}^{a\m}\p_{\n}\p_{\rho}\tilde{h}^{\rho}_{a}F_1(\Box)\p_{\m}\p_{\lambda}\tilde{h}^{\lambda\n}-2\tilde{h}^{a\m}\p_{\n}\p_{\rho}\tilde{h}^{\rho}_{a}F_1(\Box)\Box\tilde{h}^{\n}_{\m}\\
&-2\tilde{h}^{a\m}\p_{\n}\p_{\rho}\tilde{h}^{\rho}_{a}F_1(\Box)\p^{\n}\p_{\m}\tilde{h}+2\tilde{h}^{a\m}\p_{\n}\p_{\rho}\tilde{h}^{\rho}_{a}F_1(\Box)\p_{\m}\p_{\lambda}\tilde{h}^{\lambda\n}+2\tilde{h}_{cd}\p_a\p_{\n}\tilde{h}^{cd}F_1(\Box)\Box\tilde{h}^{a\n}-2\tilde{h}^{\rho\lambda}\p_{\rho}\p_a\tilde{h}_{\lambda\n}F_1(\Box)\Box\tilde{h}^{a\n}\\
&-2\tilde{h}^{\rho\lambda}\p_{\rho}\p_{\n}\tilde{h}_{\lambda a}F_1(\Box)\Box\tilde{h}^{a\n}+2\tilde{h}^{\rho\lambda}\p_{\rho}\p_{\lambda}\tilde{h}_{ a\n}F_1(\Box)\Box\tilde{h}^{a\n}+2\tilde{h}_{cd}\p_a\p_{\n}\tilde{h}^{cd}F_1(\Box)\p^{\n}\p^{a}\tilde{h}-2\tilde{h}^{\rho\lambda}\p_{\rho}\p_{a}\tilde{h}_{\lambda \n}F_1(\Box)\p^{\n}\p^{a}\tilde{h}-2   \tilde{h}^{\rho\lambda}\p_{\rho}\p_{\n}\tilde{h}_{\lambda a}F_1(\Box)\p^{\n}\p^{a}\tilde{h}+2\tilde{h}^{\rho\lambda}\p_{\rho}\p_{\lambda}\tilde{h}_{a \n}F_1(\Box)\p^{\n}\p^{a}\tilde{h}\\
&-2\tilde{h}^{\rho\lambda}\p_{\rho}\p_{\n}\tilde{h}_{\lambda a}F_1(\Box)\p^{\n}\p^{a}\tilde{h}+2\tilde{h}^{\rho\lambda}\p_{\rho}\p_{\lambda}\tilde{h}_{a \n}F_1(\Box)\p^{\n}\p^{a}\tilde{h}-4\tilde{h}_{cd}\p_a\p_{\n}\tilde{h}^{cd}F_1(\Box)\p^a\p_{\lambda}\tilde{h}^{\lambda\n}\\
&+8\tilde{h}^{\rho\lambda}\p_{\rho}\p_{a}\tilde{h}_{\lambda\n}F_1(\Box)\p_{c}\p^{(a}\tilde{h}^{\n)c}-4\tilde{h}^{\rho\lambda}\p_{\rho}\p_{\lambda}\tilde{h}_{a\n}F_1(\Box)\p^a\p_{c}\tilde{h}^{c\n},
\end{split}
\ee
\be
\label{A7}
\begin{split}
&U^{\m\n aa_1a_2bb_1b_2cd}_{2\text{Nonlocal}}\p_a \tilde{h}_{\m\n}\p_b\tilde{h}_{a_1a_2}F_1(\Box)\p_c\p_d\tilde{h}_{b_1b_2}=3\p_{\al}\tilde{h}_{\m\n}\p^{\al}\tilde{h}^{\m\n}F_1(\Box)\p_{a}\p_{b}\tilde{h}^{ab}-4\p_a\tilde{h}^a_{\rho}\p_b\tilde{h}^{\rho b}F_{1}(\Box)\p_{\m}\p_{\n}\tilde{h}^{\m\n}\\
&+4\p_a\tilde{h}^a_{\rho}\p^{\rho}\tilde{h}F_1(\Box)\p_{\m}\p_{\n}\tilde{h}^{\m\n}-2\p_{\m}\tilde{h}_{\n\al}\p^{\al}\tilde{h}^{\m\n}F_1(\Box)\p_{a}\p_{b}\tilde{h}^{ab}-\p_{\rho}\tilde{h}\p^{\rho}\tilde{h}F_1(\Box)\p_{\m}\p_{\n}\tilde{h}^{\m\n}+\p_{\m}\tilde{h}\p^{\m}\tilde{h}F_1(\Box)\Box\tilde{h}\\
&-3\p_{\al}\tilde{h}_{\m\n}\p^{\al}\tilde{h}^{\m\n}F_1(\Box)\Box\tilde{h}+4\p_a\tilde{h}^{a}_{\m}\p_b\tilde{h}^{\m b}F_1(\Box)\Box    \tilde{h}-4\p_a\tilde{h}^a_{\m}\p^{\m}\tilde{h}F_1(\Box)\Box\tilde{h}+2\p_{\m}\tilde{h}_{\n\al}\p^{\al}\tilde{h}^{\m\n}F_1(\Box)\Box\tilde{h}\\
&+2\p^{\lambda}\tilde{h}\p_a\tilde{h}_{\lambda\n}F_1(\Box)\Box\tilde{h}^{a\n}-\p^{\lambda}\tilde{h}\p_{\lambda}\tilde{h}_{a\n}F_1(\Box)\Box\tilde{h}^{a\n}-\p_{a}\tilde{h}^{\rho\lambda}\p_{\n}\tilde{h}_{\rho\lambda}F_1(\Box)\Box\tilde{h}^{a\n}-\p^{\lambda}\tilde{h}^{\rho}_{a}\p_{\rho}\tilde{h}_{\lambda\n}F_1(\Box)\Box\tilde{h}^{a\n}\\
&+\p^{\lambda}\tilde{h}^{\rho}_{a}\p_{\lambda}\tilde{h}_{\rho\n}F_1(\Box)\Box\tilde{h}^{a\n}+\p^{\rho}\tilde{h}^{\lambda}_{a}\p_{\rho}\tilde{h}_{\lambda\n}F_1(\Box)\Box\tilde{h}^{a\n}-\p^{\rho}\tilde{h}^{\lambda}_{a}\p_{\lambda}\tilde{h}_{\rho\n}F_1(\Box)\Box\tilde{h}^{a\n}+2\p^{\lambda}\tilde{h}\p_{a}\tilde{h}_{\lambda\n}F_1(\Box)\p^{\n}\p^{a}\tilde{h}\\
&- \p^{\lambda}\tilde{h}\p_{\lambda}\tilde{h}_{a\n}F_1(\Box)\p^{\n}\p^{a}\tilde{h}-  \p_a\tilde{h}^{\rho\lambda}\p_{\n}\tilde{h}_{\rho\lambda}   F_1(\Box)\p^{\n}\p^{a}\tilde{h}     -\p^{\lambda}\tilde{h}^{\rho}_a\p_{\rho}\tilde{h}_{\lambda\n} F_1(\Box)\p^{\n}\p^{a}\tilde{h}+\p^{\lambda}\tilde{h}^{\rho}_a\p_{\lambda}\tilde{h}_{\rho\n} F_1(\Box)\p^{\n}\p^{a}\tilde{h}\\
&+\p^{\rho}\tilde{h}^{\lambda}_a\p_{\rho}\tilde{h}_{\lambda\n} F_1(\Box)\p^{\n}\p^{a}\tilde{h}-\p^{\rho}\tilde{h}^{\lambda}_a\p_{\lambda}\tilde{h}_{\rho\n} F_1(\Box)\p^{\n}\p^{a}\tilde{h}-4\p^{\lambda}\tilde{h}\p_a\tilde{h}_{\lambda\n}F_1(\Box)\p_{c}\p^{(a}\tilde{h}^{\n) c}+2\p^{\lambda}\tilde{h}\p_{\lambda}\tilde{h}_{a\n}F_1(\Box)\p^a\p_c\tilde{h}^{c\n}\\
&+2\p_{a}\tilde{h}^{\rho\lambda}\p_{\n}\tilde{h}_{\rho\lambda}F_1(\Box)\p_{c}\p^{(a}\tilde{h}^{\n) c}+2\p^{\lambda}\tilde{h}^{\rho }_a\p_{\rho}\tilde{h}_{\n\lambda}F_1(\Box)\p_{c}\p^{(a}\tilde{h}^{\n) c}-2\p^{\lambda}\tilde{h}^{\rho }_a\p_{\lambda}\tilde{h}_{\n\rho}F_1(\Box)\p_{c}\p^{(a}\tilde{h}^{\n)c}\\
&-2\p^{\rho}\tilde{h}^{\lambda }_a\p_{\rho}\tilde{h}_{\n\lambda}F_1(\Box)\p_{c}\p^{(a}\tilde{h}^{\n)c}+2\p^{\rho}\tilde{h}^{\lambda }_a\p_{\lambda}\tilde{h}_{\n\rho}F_1(\Box)\p_{c}\p^{(a}\tilde{h}^{\n)c}+2\p_a\tilde{h}_{cd}\p_{\n}\tilde{h}^{cd}F_1(\Box)\Box\tilde{h}^{a\n}\\
&-2\p_{\rho}\tilde{h}^{\rho\lambda}\p_a\tilde{h}_{\lambda\n}F_1(\Box)\Box\tilde{h}^{a\n}-2\p_{\rho}\tilde{h}^{\rho\lambda}\p_{\n}\tilde{h}_{\lambda a}F_1(\Box)\Box\tilde{h}^{a\n}+2\p_{\rho}\tilde{h}^{\rho\lambda}\p_{\lambda}\tilde{h}_{a\n}F_1(\Box)\Box\tilde{h}^{a\n}\\
&+2\p_a\tilde{h}_{cd}\p_{\n}\tilde{h}^{cd}F_1(\Box)\p^{\n}\p^{a}\tilde{h}-2\p_{\rho}\tilde{h}^{\rho\lambda}\p_a\tilde{h}_{\lambda\n}F_1(\Box)\p^{\n}\p^a\tilde{h}-2\p_{\rho}\tilde{h}^{\rho\lambda}\p_{\n}\tilde{h}_{\lambda a}F_1(\Box)\p^{\n}\p^a\tilde{h}\\
&+2\p_{\rho}\tilde{h}^{\rho\lambda}\p_{\lambda}\tilde{h}_{a\n}F_1(\Box)\p^{\n}\p^a\tilde{h}-4\p_a\tilde{h}_{cd}\p_{\n}\tilde{h}^{cd}F_1(\Box)\p_c\p^{(a}\tilde{h}^{\n)c}+4\p_{\rho}\tilde{h}^{\rho\lambda}\p_{a}\tilde{h}_{\lambda\n}F_1(\Box)\p_c\p^{(a}\tilde{h}^{\n)c}\\
&+4\p_{\rho}\tilde{h}^{\rho\lambda}\p_{\n}\tilde{h}_{\lambda a}F_1(\Box)\p_c\p^{(a}\tilde{h}^{\n)c}-4\p_{\rho}\tilde{h}^{\rho\lambda}\p_{\lambda}\tilde{h}_{a\n}F_1(\Box)\p_c\p^{(a}\tilde{h}^{\n)c}.
\end{split}
\ee
Similarly, the tensor $U^{\m\n a_1a_2abb_1b_2cd}_{1\text{Nonlocal}}$ and $U^{\m\n aa_1a_2bb_1b_2cd}_{2\text{Nonlocal}}$ also exhibit symmetry under the exchange of $ a_1a_2ab \leftrightarrow b_1b_2cd$, $\m\n a \leftrightarrow a_1a_2b$, $\m\leftrightarrow\n$, $a_1\leftrightarrow a_2$, $a\leftrightarrow b$, $b_1\leftrightarrow b_2$  and $c \leftrightarrow d$. Note that the parentheses here indicate the symmetrization operation on the tensor, denoted as $T^{(ab)}=\frac{1}{2}\(T^{ab}+T^{ba} \)$.  Following a similar approach as in the GR case, we perform the Fourier transform and ultimately derive the Feynman rules for the three-graviton vertex contribution from the nonlocal terms
\be
\label{A8}
\begin{split}
2i \kappa V^{\m\n a_1a_2b_1b_2}_{h^3\text{Nonlocal}}\(l_1,l_2,l_3\)&=2i\kappa\( U^{\m\n a_1a_2abb_1b_2cd}_{1\text{Nonlocal}}l_{2a}l_{2b}F_{1}(-l^2_3)l_{3c}l_{3d}+ U^{a_1a_2 b_1b_2ab\m\n cd}_{1\text{Nonlocal}}l_{3a}l_{3b}F_{1}(-l^2_1)l_{1c}l_{1d}\right.\\
&\left. U^{b_1b_2 \m\n aba_1a_2cd}_{1\text{Nonlocal}}l_{1a}l_{1b}F_{1}(-l^2_2)l_{2c}l_{2d}+U^{\m\n aa_1a_2bb_1b_2cd}_{2\text{Nonlocal}}l_{1a}l_{2b}F_{1}(-l^2_3)l_{3c}l_{3d}\right.\\
&\left. +U^{a_1a_2ab_1b_2b\m\n cd}_{2\text{Nonlocal}}l_{2a}l_{3b}F_{1}(-l^2_1)l_{1c}l_{1d}+U^{b_1b_2a\m\n ba_1a_2cd}_{2\text{Nonlocal}}l_{3a}l_{1b}F_{1}(-l^2_2)l_{2c}l_{2d}         \).
\end{split}
\ee
Eqs. \eqref{A4} and \eqref{A8} together contribute to the three-graviton interaction in super-renormalizable gravity, where all three gravitons are off-shell. The corresponding diagram is shown in Fig. \ref{Fig. 4}.

\section{Triangle Loop Integrals}

We now turn to the triangle integrals governing the one-loop correction, shown in Fig. \ref{Fig.2}, beginning with their form in GR. Following the definitions in Eqs. \eqref{29}-\eqref{33}, we first analyze the fundamental structure of these integrals by temporarily setting aside explicit vertex factors. The simplest and most illustrative case is the scalar triangle integral
\be
\label{B1}
\begin{split}
I^{\text{GR}}=\int\frac{\mathrm{d}^D l}{(2\pi)^D}\frac{e^{-\frac{l^2}{M^2_{\ast}}}e^{-\frac{(l+q_{\bot})^2}{M^2_{\ast}}}}{l^2(l+q_{\bot})^2[(l+k)^2-m^2+i\epsilon]},
\end{split}
\ee
where the exponential factors $e^{-l^2/M^2_*}$ and $e^{-(l+q_\bot)^2/M^2_*}$ implement the weak nonlocality of the gravitational interaction. These form factors ensure UV convergence while preserving covariance, and become crucial in the classical limit where they regulate the short-distance behavior without introducing additional poles. The integral represents the one-loop exchange of two nonlocal graviton propagators between massive scalar propagators, which reduces to the standard GR expression in the limit $M_* \to \infty$.

The classical limit enables us to reduce the complexity of the massive propagator
\be
\label{B2}
\begin{split}
\frac{1}{(l+k)^2-m^2+i\epsilon}\approx\frac{1}{2kl+i\epsilon}=\frac{1}{2ml_{||}}-\frac{i\pi}{2m}\delta{(l_{||})}.
\end{split}
\ee

By substituting this equation into the scalar triangle integral of Eq. \eqref{B1}, we can disregard the first term in Eq. \eqref{B2}. This is because the two graviton propagators are even in $l_{||}$. In the classical limit, the scalar triangle integral simplifies to
\be
\label{B3}
\begin{split}
I^{\text{GR}}=-\frac{i}{4m}N_{D-1}=-\frac{i}{4m}\int\frac{{\dif^{D-1} l_{\bot}}}{(2\pi)^{D-1}}\frac{e^{-\frac{l^2_{\bot}}{M^2_{\ast}}}e^{-\frac{\(l_{\bot}+q_{\bot}\)^2}{M^2_{\ast}}} }{l^2_{\bot}(l_{\bot}+q_{\bot})^2}.
\end{split}
\ee

The integral $N_{D-1}$ defined in Eq. \eqref{B3} plays a particularly important role, as it directly corresponds to the convolution structure analyzed in the main text. In position space, this convolution reduces to simple multiplication, significantly simplifying the interpretation of the nonlocal interaction. The tensor triangle integrals $I^{\text{GR}}_{\mu}$ and $I^{\text{GR}}_{\mu\nu}$ can be treated through the same approach, with only these two tensor structures contributing to the classical gravitational potential in the GR sector. 
\be
\1\{\begin{split}
\label{B4}
&I^{\text{GR}}_{\m}=\int\frac{{\dif^D l}}{(2\pi)^D}\frac{e^{-\frac{l^2}{M^2_{\ast}}}e^{-\frac{\(l+q_{\bot}\)^2}{M^2_{\ast}}} l_{\m}}{l^2(l+q_{\bot})^2\[(l+k)^2-m^2+i\epsilon\]},\\
&I^{\text{GR}}_{\m\n}=\int\frac{{\dif^D l}}{(2\pi)^D}\frac{e^{-\frac{l^2}{M^2_{\ast}}}e^{-\frac{\(l+q_{\bot}\)^2}{M^2_{\ast}}} l_{\m}l_{\n}}{l^2(l+q_{\bot})^2\[(l+k)^2-m^2+i\epsilon\]}.
\end{split}\2.
\ee

The tensor integrals can be solved algebraically by proposing an ansatz. Let us first consider $I^{\text{GR}}_{\m}$, which can be expressed as $Aq_{\bot\m}+Bk_{\m}$. The coefficients $A$ and $B$ are determined by the equations $q^{\mu}_{\bot} I_{\mu} = q^2_{\bot} A $ and $k^{\mu} I_{\mu} = m^2 B$. Subsequently, we can apply relations like $2k^\mu l_\mu = (l + k)^2 - m^2 - l^2 $ to simplify the numerators, reducing them to scalar integrals that no longer contain loop momenta in the numerator. It can be shown that $B=\frac{iN_{D}}{2m^2}$, with $N_D$ representing the $D$-dimensional integral
\be
\label{B5}
\begin{split}
N_{D}=\int\frac{{\dif^D l_{E}}}{(2\pi)^D}\frac{e^{\frac{l^2_{E}}{M^2_{\ast}}}e^{\frac{\(l_E+q_{\bot}\)^2}{M^2_{\ast}}} }{l^2_E(l_E+q_{\bot})^2},
\end{split}
\ee
where the integral discussed above belong to the category of nonlocal integrals, making the conventional Wick rotation in QFT inapplicable. However, an alternative method can be employed to handle such integral \cite{Buoninfante:2022krn}. Specifically, the integration variable $l_0$ can be defined on the imaginary axis, which is equivalent to condition $l_0=il_D$ ($l^2=-l^2_E$). Subsequently, the scattering amplitude calculated on the imaginary axis is mapped to the physical conclusion on the real axis.

Additionally, we can also conclude that $A=\frac{iN_{D-1}}{8m}$. Since this is a spatial integral, no above integration steps are necessary. In particular, we neglect $\int\frac{{\dif^{D-1} l_{\bot}}}{(2\pi)^{D-1}}\frac{e^{-\frac{l^2_{\bot}}{M^2_{\ast}}}e^{-\frac{\(l_{\bot}+q_{\bot}\)^2}{M^2_{\ast}}} }{l^2_{\bot}}$, $\int\frac{{\dif^{D-1} l_{\bot}}}{(2\pi)^{D-1}}\frac{e^{-\frac{l^2_{\bot}}{M^2_{\ast}}}e^{-\frac{\(l_{\bot}+q_{\bot}\)^2}{M^2_{\ast}}} }{(l_{\bot}+q_{\bot})^2}$ in the proof, as they do not contribute non-analytic terms in the classical limit. The final expression for $I^{\text{GR}}_{\m}$ becomes
\be
\label{B6}
\begin{split}
I^{\text{GR}}_{\m}=\frac{iN_{D-1}}{8m}q_{\bot\m}+\frac{i N_{D}}{2m^2} k_{\m}.
\end{split}
\ee
Ultimately, we apply a similar approach to derive
\be
\label{B7}
\begin{split}
I^{\text{GR}}_{\m\n}=-\frac{i q^2_{\bot}N_{D-1}}{16m(D-2)}\[\(D-1\)\frac{q_{\bot\m}q_{\bot\n}}{q^2_{\bot}}-\eta^{\bot}_{\m\n} \]-\frac{i N_{D}}{4m^2} \(   k_{\m}q_{\bot\n}+k_{\n}q_{\bot\m} \).
\end{split}
\ee

Now, we focus on the integrals associated with the contributions from nonlocal terms, which are categorized into three types as defined in Eq. \eqref{33}. Using the same approach, we directly present the final results for the integrals involved.

The integral of first type
\be
\label{B8}
\begin{split}
&I^{(1)}_{\m}=\frac{i N^{(1)}_{D-1}}{8m}q_{\bot \m}+\frac{i N^{(1)}_{D}}{2m^2} k_{\m},\\
&I^{(1)}_{\m\n}=-\frac{i q^2_{\bot}N^{(1)}_{D-1}}{16m(D-2)}\[\(D-1\)\frac{q_{\bot\m}q_{\bot\n}}{q^2_{\bot}}-\eta^{\bot}_{\m\n} \]-\frac{i N^{(1)}_{D}}{4m^2} \(   k_{\m}q_{\bot\n}+k_{\n}q_{\bot\m} \),\\
&I^{(1)}_{\m\n\rho}=\frac{i (D+1)N^{(1)}_{D-1}}{32m(D-2)}q_{\bot\m}q_{\bot\n}q_{\bot\rho}-\frac{i q^2_{\bot}N^{(1)}_{D-1}}{32m(D-2)}\(q_{\bot\m}\eta^{\bot}_{\n\rho}+q_{\bot\n}\eta^{\bot}_{\m\rho}+q_{\bot\rho}\eta^{\bot}_{\m\n} \)\\
&-\frac{i q^2_{\bot}N^{(1)}_{D}}{8m^4(D-1)}k_{\m}k_{\n}k_{\rho}+\frac{i D N^{(1)}_{D}}{8m^2(D-1)}\( k_{\m}q_{\bot\n}q_{\bot\rho} +k_{\n}q_{\bot\m}q_{\bot\rho}+k_{\rho}q_{\bot\m}q_{\bot\n} \)\\
&-\frac{i q^2_{\bot}N^{(1)}_{D} }{8m^2(D-1)}\( k_{\m}\eta^{\bot}_{\n\rho}+k_{\n}\eta^{\bot}_{\m\rho}+k_{\rho}\eta^{\bot}_{\m\n}  \),  \\
&I^{(1)}_{\m\n\rho\sigma}= -\frac{i}{4m}\[ \frac{(D+1)(D+3)N^{(1)}_{D-1}}{16D(D-2)}q_{\bot\m}q_{\bot\n}q_{\bot\rho}q_{\bot\sigma}+\frac{q^4_{\bot}N^{(1)}_{D-1}}{16D(D-2)}\( \eta^{\bot}_{\m\n}\eta^{\bot}_{\rho\sigma}+\eta^{\bot}_{\m\rho}\eta^{\bot}_{\n\sigma}+\eta^{\bot}_{\m\sigma}\eta^{\bot}_{\n\rho} \) \right.\\
&\left. -\frac{(D+1)q^2_{\bot}N^{(1)}_{D-1}}{16D(D-2)}\(q_{\bot\m}q_{\bot\n}\eta^{\bot}_{\rho\sigma}+q_{\bot\m}q_{\bot\sigma}\eta^{\bot}_{\rho\n} +q_{\bot\m}q_{\bot\rho}\eta^{\bot}_{\n\sigma} +q_{\bot\rho}q_{\bot\sigma}\eta^{\bot}_{\m\n} +q_{\bot\n}q_{\bot\sigma}\eta^{\bot}_{\rho\m} +q_{\bot\n}q_{\bot\rho}\eta^{\bot}_{\m\sigma}   \)  \] \\
& -\frac{iq^2_{\bot}N^{(1)}_{D}}{8m^4(D-1)}\( k_{\m}k_{\n}k_{\rho}q_{\bot\sigma}+ k_{\sigma}k_{\m}k_{\n}q_{\bot\rho}+ k_{\rho}k_{\sigma}k_{\m}q_{\bot\n}+ k_{\n}k_{\rho}k_{\sigma}q_{\bot\m} \)\\
& -\frac{i(D+2)N^{(1)}_{D}}{16m^2(D-1)}\( k_{\m}q_{\bot\n}q_{\bot\rho}q_{\bot\sigma}+k_{\n}q_{\bot\m}q_{\bot\rho}q_{\bot\sigma}+k_{\rho}q_{\bot\n}q_{\bot\m}q_{\bot\sigma}+k_{\sigma}q_{\bot\n}q_{\bot\rho}q_{\bot\m} \)\\
&+\frac{i q^2_{\bot}N^{(1)}_{D-1}}{16m^2(D-1)}\( k_{\m}\eta_{\n\rho}q_{\bot\sigma} +k_{\m}\eta_{\sigma\n}q_{\bot\rho} +k_{\m}\eta_{\rho\sigma}q_{\bot\n}    +k_{\n}\eta_{\m\rho}q_{\bot\sigma}   +k_{\n}\eta_{\sigma\m}q_{\bot\rho}+k_{\n}\eta_{\sigma\rho}q_{\bot\m} \right.\\
&\left.+ k_{\rho}\eta_{\n\m}q_{\bot\sigma}+ k_{\rho}\eta_{\sigma\m}q_{\bot\n}+k_{\rho}\eta_{\n\sigma}q_{\bot\m}+k_{\sigma}\eta_{\n\m}q_{\bot\rho}+k_{\sigma}\eta_{\m\rho}q_{\bot\n}+k_{\sigma}\eta_{\n\rho}q_{\bot\m} \).
\end{split}
\ee

The integral of second type
\be
\label{B10}
\begin{split}
&I^{(2)}=-\frac{i}{4m}N^{(2)}_{D-1},\\
&I^{(2)}_{\m}=\frac{i N^{(2)}_{D-1}}{8m}q_{\bot \m}+\frac{i N^{(2)}_{D}}{2m^2} k_{\m},\\
&I^{(2)}_{\m\n}=-\frac{i q^2_{\bot}N^{(2)}_{D-1}}{16m(D-2)}\[\(D-1\)\frac{q_{\bot\m}q_{\bot\n}}{q^2_{\bot}}-\eta^{\bot}_{\m\n} \]-\frac{i N^{(2)}_{D}}{4m^2} \(   k_{\m}q_{\bot\n}+k_{\n}q_{\bot\m} \).
\end{split}
\ee

The integral of third type
\be
\label{B9}
\begin{split}
&I^{(3)}_{\m\n}=-\frac{i q^2_{\bot}N^{(3)}_{D-1}}{16m(D-2)}\[\(D-1\)\frac{q_{\bot\m}q_{\bot\n}}{q^2_{\bot}}-\eta^{\bot}_{\m\n} \]-\frac{i N^{(3)}_{D}}{4m^2} \(   k_{\m}q_{\bot\n}+k_{\n}q_{\bot\m} \),\\
&I^{(3)}_{\m\n\rho}=\frac{i (D+1)N^{(3)}_{D-1}}{32m(D-2)}q_{\bot\m}q_{\bot\n}q_{\bot\rho}-\frac{i q^2_{\bot}N^{(3)}_{D-1}}{32m(D-2)}\(q_{\bot\m}\eta^{\bot}_{\n\rho}+q_{\bot\n}\eta^{\bot}_{\m\rho}+q_{\bot\rho}\eta^{\bot}_{\m\n} \)\\
&-\frac{i q^2_{\bot}N^{(3)}_{D}}{8m^4(D-1)}k_{\m}k_{\n}k_{\rho}+\frac{i D N^{(3)}_{D}}{8m^2(D-1)}\( k_{\m}q_{\bot\n}q_{\bot\rho} +k_{\n}q_{\bot\m}q_{\bot\rho}+k_{\rho}q_{\bot\m}q_{\bot\n} \)\\
&-\frac{i q^2_{\bot}N^{(3)}_{D} }{8m^2(D-1)}\( k_{\m}\eta^{\bot}_{\n\rho}+k_{\n}\eta^{\bot}_{\m\rho}+k_{\rho}\eta^{\bot}_{\m\n}  \). \\
\end{split}
\ee

In the above expression, six integrals are present, three of which contribute to the Newtonian potential. To facilitate numerical evaluation, we can represent them as follows
\be
\label{B11}
\begin{split}
&N^{(1)}_{D}=\int\frac{{\dif^D l_{E}}}{(2\pi)^D}\frac{e^{\frac{l^2_{E}}{M^2_{\ast}}}e^{\frac{\(l_E+q_{\bot}\)^2}{M^2_{\ast}}}\(e^{\frac{-\(l_E+q_{\bot}\)^2}{M^2_{\ast}}}-1\) }{l^2_E(l_E+q_{\bot})^4},\\
&N^{(2)}_{D}=-\int\frac{{\dif^D l_{E}}}{(2\pi)^D}\frac{e^{\frac{l^2_{E}}{M^2_{\ast}}}e^{\frac{\(l_E+q_{\bot}\)^2}{M^2_{\ast}}}\( e^{\frac{q^2_{\bot}}{M^2_{\ast}}}-1 \) }{l^2_Eq^2_{\bot}(l_E+q_{\bot})^2},\\
&N^{(3)}_{D}=\int\frac{{\dif^D l_{E}}}{(2\pi)^D}\frac{e^{\frac{l^2_{E}}{M^2_{\ast}}}e^{\frac{\(l_E+q_{\bot}\)^2}{M^2_{\ast}}} \(e^{\frac{-l^2_E}{M^2_{\ast}}}-1\) }{l^4_E(l_E+q_{\bot})^2},
\end{split}
\ee
and
\be
\label{B12}
\begin{split}
&N^{(1)}_{D-1}=-\int\frac{{\dif^{D-1} l_{\bot}}}{(2\pi)^{D-1}}\frac{e^{-\frac{l^2_{\bot}}{M^2_{\ast}}}e^{-\frac{\(l_{\bot}+q_{\bot}\)^2}{M^2_{\ast}}}\( e^{\frac{\(l_{\bot}+q_{\bot}\)^2}{M^2_{\ast}}}-1  \)       }{l^2_{\bot}(l_{\bot}+q_{\bot})^4},\\
&N^{(2)}_{D-1}=-\int\frac{{\dif^{D-1} l_{\bot}}}{(2\pi)^{D-1}}\frac{e^{-\frac{l^2_{\bot}}{M^2_{\ast}}}e^{-\frac{\(l_{\bot}+q_{\bot}\)^2}{M^2_{\ast}}} \( e^{\frac{q^2_{\bot}}{M^2_{\ast}}}-1\)      }{l^2_{\bot}q^2_{\bot}(l_{\bot}+q_{\bot})^2},\\
&N^{(3)}_{D-1}=-\int\frac{{\dif^{D-1} l_{\bot}}}{(2\pi)^{D-1}}\frac{e^{-\frac{l^2_{\bot}}{M^2_{\ast}}}e^{-\frac{\(l_{\bot}+q_{\bot}\)^2}{M^2_{\ast}}}  \(e^{\frac{l^2_{\bot}}{M^2_{\ast}}}-1\)       }{l^4_{\bot}(l_{\bot}+q_{\bot})^2}.
\end{split}
\ee

\bibliographystyle{unsrt}
\bibliography{AmpNQG}
\end{document}